\documentclass[final,5p,times,twocolumn,authoryear]{elsarticle}
\usepackage{amssymb}
\usepackage{amsmath}
\usepackage[caption=true]{subfig}
\usepackage{multirow}
\usepackage{algorithm}
\usepackage{enumerate}
\usepackage{algpseudocode}
\usepackage{float}
\usepackage[colorlinks=true,linkcolor=black, citecolor=blue, urlcolor=blue]{hyperref}
\usepackage{tikz,pgfplots}
\usepackage{breqn}
\usepackage{widetext}
\usepackage{bbm}
\usepackage{cleveref}
\usepackage{mathtools}
\usepackage{stackengine}
\usepackage{widetext}
\pgfplotsset{%
    compat=newest, 
    tick label style={font=\footnotesize},
    label style={font=\small},
    legend style={font=\small},
    axis x line = center,
    axis y line = center,
    every axis/.style={pin distance=1ex},
    trim axis left
    } 

\journal{arXiv}

\makeatletter
\def\ps@pprintTitle{%
  \let\@oddhead\@empty
  \let\@evenhead\@empty
  \def\@oddfoot{\reset@font\hfil\thepage\hfil}
  \let\@evenfoot\@oddfoot
}
\makeatother
\begin{document}
\begin{frontmatter}

\title{Detecting discrete processes with the Epps effect}

\author[uct-sta,ox-eng]{Patrick Chang}
\ead{patrick.chang@maths.ox.ac.uk}
\author[uct-sta]{Etienne Pienaar}
\ead{etienne.pienaar@uct.ac.za}
\author[uct-sta]{Tim Gebbie}
\ead{tim.gebbie@uct.ac.za}
\address[uct-sta]{Department of Statistical Science, University of Cape Town, Rondebosch 7700, South Africa}
\address[ox-eng]{Mathematical Institute, University of Oxford, University of Oxford, Oxford OX1 3PJ, United Kingdom}

\begin{abstract}
The Epps effect is key phenomenology relating to high frequency correlation dynamics in financial markets. We argue that it can be used to provide insight into whether tick data is best represented as samples from Brownian diffusions, or as samples from truly discrete events represented as connected point processes. We derive the Epps effect arising from asynchrony and provide a refined method to correct for the effect. We then propose three experiments which show how to discriminate between possible underlying representations. These in turn demonstrate how a simple Hawkes representation recovers phenomenology reported in the literature that cannot be recovered using a Brownian representation without additional {\it ad hoc} model complexity. However, complex {\it ad hoc} noise models built on Brownian motions cannot in general be discriminated relative to a Hawkes representation. Nevertheless, we argue that high frequency correlation dynamics are most faithfully recovered when tick data is represented as a web of interconnected discrete events rather than being samples from continuous Brownian diffusions even when combined with noise.
\end{abstract}


\end{frontmatter}


\begin{abstract}
The Epps effect is key phenomenology relating to high frequency correlation dynamics in financial markets. We argue that it can be used to provide insight into whether tick data is best represented as samples from Brownian motions (diffusions), or as samples from truly discrete events represented as connected point processes. We derive the Epps effect arising from asynchrony and provide a refined method to correct for the effect. We then propose three experiments which show how to discriminate between possible representations of the underlying process. These in turn demonstrate how a simple Hawkes representation recovers phenomenology reported in the literature that cannot be recovered using a Brownian representation without the addition of {\it ad hoc} model complexity. However, complex {\it ad hoc} noise models built on Brownian motions cannot in general be discriminated relative to a Hawkes representation. 
\end{abstract}


\section{Introduction}\label{sec:intro}

Models are abstractions which aim to faithfully recover, explain, and sometimes predict what we observe in real-world data generating processes. When proposing model structures for such processes, we make explicit assumptions about the nature of the underlying process. Depending on the mathematical machinery we use to build the model process, we may also incur implicit assumptions that depart from the observed reality. Here we are concerned with such a departure observed in the application of diffusion models to high frequency financial data. That is, the decay of correlations as the lengths of sampling intervals decrease in high frequency financial markets---the Epps effect \cite{EPPS1979}. Models where the underlying representation is cast in terms of Brownian motions, so-called diffusion models, implicitly assume that correlations exist at infinitesimal scales --- an implicit consequence of including correlated Brownian motions in the diffusion model equations. 

In the real-world where processes can only be observed discretely, and on finite time scales, if such a model is a plausible representation of the true data generating process, one should expect that empirical correlations can be recovered at all sampling scales---provided a sufficiently long series is available. Empirically, the Epps effect constitutes a departure of the model process from reality, which needs to be corrected for.

In contrast, the fine-to-coarse model proposed by \citet{BDHM2013a} using a mutually exciting Hawkes process implicitly assumes that correlations depend on the sampling interval $\Delta t$. Therefore, under such a model, the Epps effect is a genuine effect and we only need to correct for statistical causes leading to an additional decay in correlations. That is, departures from the decay in correlations over and above what is expected under the model process. 

Although it is difficult to determine what truly constitutes a {\it good} model in the present context, one may still determine the usefulness of a model in its ability to recover certain stylised facts and interrogate these by obtaining results from simulations that align with what is observed empirically. In our case, a good model should be able to recover the Epps effect in its entirety. This has been problematic in the literature where the underlying representation or model data generating process is based purely on Brownian motions \cite{MSG2011}. Such models only explain a fraction of the empirically observed Epps effect and thus additional {\it ad hoc} complexity such as market microstructure noise have been included in the model specification to try and conform such models with observations. These are latent models where the observed process consists of a diffusion process with noise added to the trajectory, with the diffusion thus being treated as a latent component \cite{ZHANG2010,RobertChristianY.2011Anaf,AMZ2011}. Our concern with these specifications is that it may conceal an inappropriate representation of the underlying process.

The Epps effect is pervasive and has been observed in stock markets and foreign exchange markets \cite{MMZ2011}. The literature has isolated three primary potential sources: (i) asynchrony, (ii) lead-lag and (iii) tick-size. Concretely, \citet{RENO2001} explored the effect of asynchrony in the presence of lead-lag effects. \citet{PI2007} demonstrated the impact different levels of asynchrony has on the Epps effect. Around the same time, \citet{TK2007} derived an analytical expression characterising the Epps effect as a function of the rate parameter under Poisson sampling. This lead to the realisation that one can decompose the correlation measured at time intervals of length $\Delta t$ as a function of the correlation at smaller time intervals of length $\Delta t_0$ \cite{TK2009}. Subsequently, \citet{MSG2010} investigated the direct impact of tick-size on the Epps effect. They were able to find a representation that combined the compensation of tick-size with asynchrony \cite{MSG2011}. The analytical expression characterising the Epps effect in \citet{TK2007} was then further extended to separate the effects from asynchrony and that of lead-lag \cite{MMZ2011}. More recently, \citet{PCEPTGAS} investigated the Epps effect under varying definitions of time.

Asynchrony is a statistical cause contributing to a decay in correlations as the sampling interval decreases. Many methods have been proposed to fix or compensate for this. The now canonical approach of \citet{HY2005} proposed a cumulative covariance estimator. \citet{BLPP2019} characterised the Epps effect in terms of the probability of flat trading and provided a correction for the effect. Here we provide a correction based on the distortion resulting from partial overlapping intervals from the sampling process \cite{TK2007,MMZ2011,MSG2011}. Methods aimed at compensating for asynchrony as a cause of the effect all assume that there is some underlying synchronous process with some true correlation structure. Our concern is with the underlying representation of the synchronous process itself when there are no true correlations. 

We demonstrate that the {\it residual Epps effect}, the remaining Epps effect after correcting for statistical causes, allows us to discriminate between a diffusion-based representation for the underlying process and a diffusion-based process with {\it ad hoc} noise or discrete connected events. Currently we cannot discriminate between diffusion-based processes with {\it ad hoc} noise and processes from discrete connected events. This is because both configurations lead to similar behaviour in the residual Epps effect, albeit for completely different reasons. 

Finding the correct representation is important, and we argue that the Hawkes representation is likely more appropriate since it does not require {\it ad hoc} model complexity to recover the observed phenomenology. The effectiveness and simplicity of the Hawkes representation draws attention to the primacy of the waiting time process as an appropriate description of real-world financial processes at the atomic level; The time between events is a fundamental property and from this property (with appropriate averaging, and limits) the observed properties of financial market data can be retrieved under the model process.  

Outside of the Epps literature there have been a large number of studies concerning irregularly spaced data \cite{RobertChristianY.2011Anaf}. Here these have arisen from two important strands of the literature: duration models \cite{EngleRussell1998}, and intensity models \cite{BOWSHER2007}. Much of the mathematical finance literature starts with continuous Ito semi-martingales, even when it is known that real financial market data does not conform to this representation without significant adjustments to the model.

In the idealistic world of continuous time representations, it can be argued that prices are only observed when markets trade and that this price is sampled from some continuous underlying process \cite{Robert2011, RobertChristianY.2011Anaf}. Such models introduce a sampling process but retain the underlying Brownian latent model. It must be noted that there are in fact classes of models that can generate correlations as emergent properties extrinsic to the discrete data generating processes \cite{bauer2024correlation,DIANA2025116202}. Here models are phenomenological rather than based on a foundational stochastic process representation, and the discrete model can be averaged to match some continuous-time semi-martingale representation in an appropriate diffusion limit \cite{DIANA2025116202}. 

The argument put forward here is that if a foundational stochastic process representation is to be retained, in order to retain parsimony it should start with the waiting time distributions between events, then include some averaging process; and should not start with some continuous-time semi-martingale based latent model and then engage in a combination of sampling and the inclusion of {\it ad hoc} noise models. Simple because we observe the prior in real financial markets, and not the latter.

The paper proceeds as follows. Section~\ref{sec:Epps} demonstrates how the Epps effect arises from asynchrony. We introduce three methods to compensate for this effect. Section~\ref{sec:experiments} compares the various corrections for asynchrony under different underlying representations. Section~\ref{sec:key} demonstrates how we can use the residual Epps effect to discriminate the underlying process. Section~\ref{sec:NoiseEst} briefly provides some alternative estimation methods and discusses their strengths and weaknesses. Section~\ref{sec:noise} discusses the limitations of our discrimination and the importance of finding the correct representation. Section~\ref{sec:emp} we discriminate the underlying representation on trade and quote data from the Johannesburg Stock Exchange (JSE). Finally, we end with some closing remarks in Section~\ref{sec:conclusion}.

\section{Asynchrony}\label{sec:Epps}

\subsection{The Epps effect from asynchrony}\label{subsec:inducing}

Consider a multivariate diffusion process as the underlying representation of the stochastic nature of a log-price of the $i$th asset at time $t$. Assuming that the process has stationary increments $dX_t^i$, let the finite variation over the interval $\Delta t$ be given by
$$
X_{\Delta t}^i = \int_0^{\Delta t} dX_t^i.
$$
The infinitesimal lagged correlation between assets $i$ and $j$ is taken to be
\begin{equation}\label{epps:eq:2}
    \left< dX_t^i dX_{t'}^j \right> = c_{t-t'}^{ij} dt dt',
\end{equation}
where $c_{t-t'}^{ij}$ is defined as
$$
c_{t-t'}^{ij}
    =\begin{cases} 
      \delta_{t-t'} & \text{if } i = j, \\
      c \delta_{t-t'} & \text{if } i \neq j,
    \end{cases}
$$
and $\delta_x$ denotes the delta function (unit impulse) at 0. This means $\delta_x$ is equal to 1 when $x=0$ and is equal 0 when $x \neq 0$. We want to examine the behaviour of the covariance process $C^{ij}_{\Delta t} = \left< X_{\Delta t}^i X_{\Delta t}^j \right>$ on time scales $\Delta t$. This can be computed using Eq. \eqref{epps:eq:2} as
$$
C^{ij}_{\Delta t} = \int_0^{\Delta t} \int_0^{\Delta t} c_{t-t'}^{ij} dt dt'.
$$
Following the double integration and unit impulse, it follows that for the synchronous case we have the variance and covariance as $C^{ii}_{\Delta t} = \Delta t$ and $C^{ij}_{\Delta t} = c \Delta t$, respectively.
Thus the correlation of the synchronous process is
$$
\rho_{\Delta t}^{ij} = \frac{C^{ij}_{\Delta t}}{\sqrt{C^{ii}_{\Delta t}  \cdot C^{jj}_{\Delta t}}} = c.
$$
We see that for the synchronous process both the variance and covariance scale linearly with the sampling interval $\Delta t$. Thus the correlation is independent of $\Delta t$. \cite{MMZ2011} characterize the Epps effect by saying that the Epps effect is present whenever $\rho_{\Delta t}^{ij}$ depends on $\Delta t$ and is absent otherwise.

\begin{figure}[tb]
    \centering
    \includegraphics[width=0.5\textwidth]{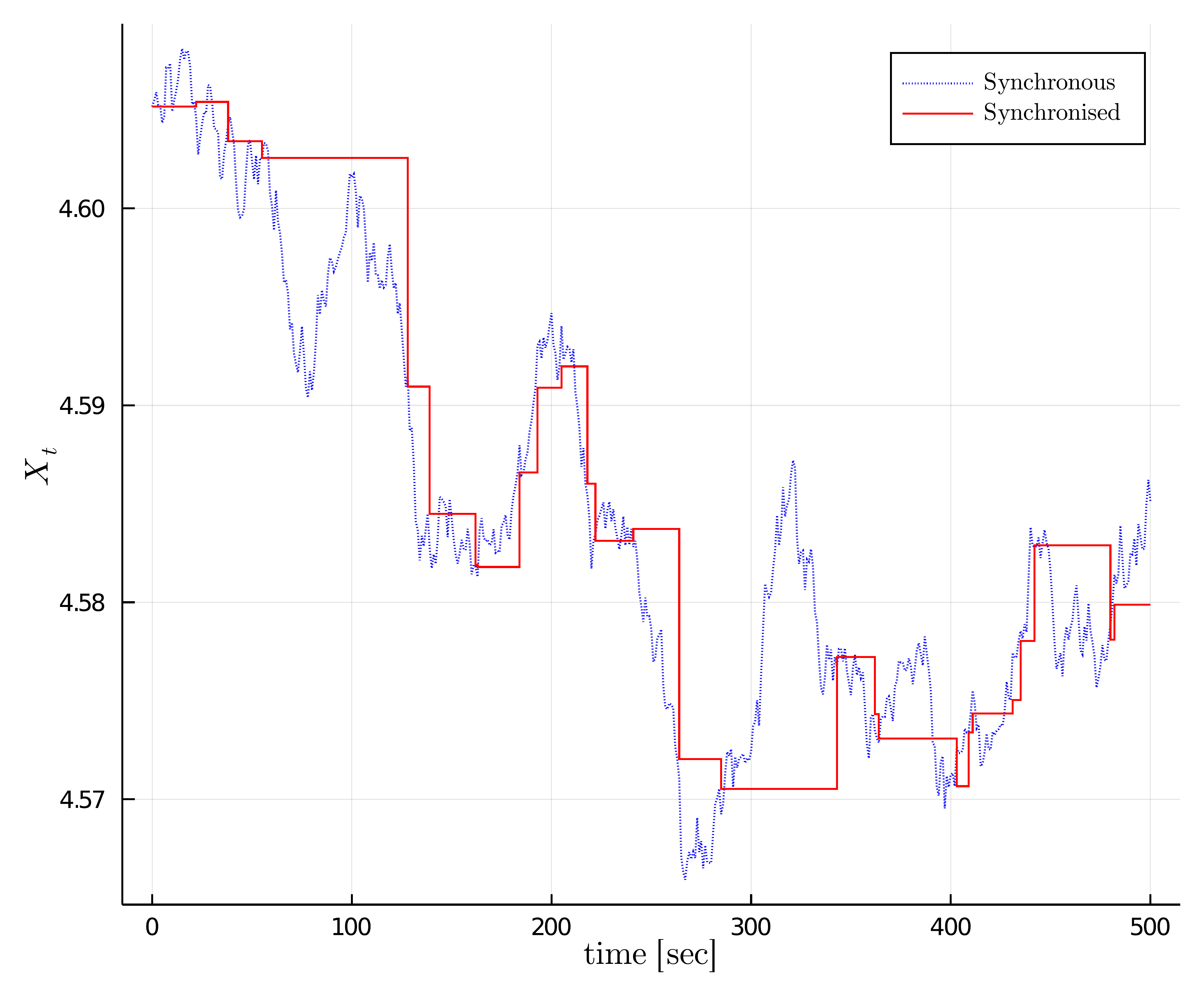}
    \caption{A comparison of a realisation of the synchronous process $X_{\Delta t}^i$ (dotted line) and the synchronised process $\tilde{X}_{\Delta t}^i$ (solid line). Here the inter-arrival distribution of $U^i$ follows an exponential with $\lambda = 1/15$.}
    \label{fig:GBMPricePaths}
\end{figure}

A property of high frequency finance is that tick-by-tick trades arrive in an asynchronous fashion following an arrival type representation where the inter-arrivals are not equidistant. The Epps effect arising from asynchrony is derived by synchronising the asynchronous observations using previous tick interpolation. To this end, let $U^i = \{ t_k^i \}_{k \in \mathbb{Z}}$ be the set of asynchronous arrival times observed in the interval $\left[0, T\right]$ for a realisation of the underlying synchronous process $X_t^i$. Using these two components we can construct the synchronised process
$$
\tilde{X}_{\Delta t}^i = \int_{t_1^i}^{t_2^i} dX_t^i,
$$
where $t_1^i = \max \{t_k^i \in U^i | t_k^i \leq 0 \}$ and $t_2^i = \max \{t_k^i \in U^i | t_k^i \leq \Delta t \}$. The resulting synchronised process is piece-wise constant with jumps at $t_k^i \in U^i$. A comparison of the two process $X_{\Delta t}^i$ and $\tilde{X}_{\Delta t}^i$ can be seen in Figure~\ref{fig:GBMPricePaths}.

Given $U^i$ and $U^j$, the covariance of the synchronised process can be defined as
\begin{equation}\label{epps:eq:8}
    \begin{aligned}
        \tilde{C}_{\Delta t}^{ij} 
        = \mathbb{E}\left[ \left< \tilde{X}_{\Delta t}^i \tilde{X}_{\Delta t}^j \right> \middle\vert {{U^i},{U^j}} \right] = \mathbb{E}\left[ \int_{t_1^i}^{t_2^i} \int_{t_1^j}^{t_2^j} c_{t-t'}^{ij} dt dt' \middle\vert {{U^i},{U^j}} \right] \nonumber \\
        \end{aligned}
        \end{equation}
        Here $\mathbb{E} \left[ \cdot \right]$ is the expectation of the sampling process. The covariance is then
\begin{equation} \label{epps:eq:8.1}
        \tilde{C}_{\Delta t}^{ij}  = c \cdot \mathbb{E} \left[ \left\lvert \left[t_1^i, t_2^i\right] \cap \left[t_1^j, t_2^j\right] \right\rvert \right].
\end{equation}
Similarly, the variance is
$$
\begin{aligned}
        \tilde{C}_{\Delta t}^{ii} = \mathbb{E} \left[ \left\lvert \left[t_1^i, t_2^i\right] \cap \left[t_1^i, t_2^i\right] \right\rvert \right].
\end{aligned}
$$
Given $U^i$ and $U^j$, we can estimate the expectation of the overlap of the sampling process at a particular discretisation size $\Delta t$ by defining a new variable $\gamma_i(t) = \max \{t_k^i : t_k^i \leq t \}$ for $t \in \left[0,T\right]$. The expectation can then be estimated as
\begin{equation}\label{epps:eq:10}
    \hat{\kappa}_{\Delta t}^{ij} = \mathbb{E} \left[ \left\lvert \left[\gamma_i(t-\Delta t), \gamma_i(t)\right] \cap \left[\gamma_j(t-\Delta t), \gamma_j(t)\right] \right\rvert \right].
\end{equation}
Thus the correlation of the synchronised process is then given by
\begin{equation}\label{epps:eq:11}
    \tilde{\rho}_{\Delta t}^{ij} = c \cdot \frac{\hat{\kappa}_{\Delta t}^{ij}}{\sqrt{{\hat{\kappa}_{\Delta t}^{ii} \cdot \hat{\kappa}_{\Delta t}^{jj}}}}.
\end{equation}
This depends on $\Delta t$ and therefore the Epps effect is present.

The correlation from the synchronous process $X_{\Delta t}$ is being extracted when the intervals $[{t_1^i, t_2^i}]$ and $[{t_1^j, t_2^j}]$ overlap, whereas the non-overlapping intervals are uncorrelated \cite{MSG2011}. Therefore the correlation of the synchronised process will be distorted by the non-overlapping intervals.

This approach has been previously investigated \cite{TK2007, MMZ2011, MSG2011}. The key difference between the work of \citet{TK2007} and \citet{MMZ2011} is that a specific distribution for the sampling process with stationary increments is chosen beforehand, usually a homogeneous Poisson process. Here we only require that the sampling process have stationary increments. This can be useful because characterising the Epps effect with a specific {\it a-priori} distribution requires knowledge about the distribution of $\mathcal{W}_{\Delta t}$ where
$$
\mathcal{W}_{\Delta t} = \min \left\{ {\gamma_i(t), \gamma_j(t)} \right\} - \max \left\{\gamma_i(t-\Delta t), \gamma_j(t-\Delta t) \right\}.
$$
This is not always easy to obtain. In the case of Poisson sampling we can find the exact distribution of $\mathcal{W}_{\Delta t}$ which means Eq. \eqref{epps:eq:11} reduces to
\begin{equation}\label{epps:eq:12}
    \tilde{\rho}_{\Delta t}^{ij} = c \left( 1 + \frac{1}{\lambda \Delta t} \left( e^{-\lambda \Delta t} - 1 \right) \right).
\end{equation}
There is a fundamental difference between our approach and that of \citet{MSG2011}. They compensate each of the overlapping return intervals while they are estimating the (compensated) correlation. 
In contrast, we separate the estimation of the correlation $\tilde{\rho}_{\Delta t}^{ij}$ and the compensation factor $\hat{\kappa}_{\Delta t}^{ij}$ as separate quantities. This is a crucial nuance that allows us to determine both the measured correlation from the synchronised process $\tilde{\rho}_{\Delta t}^{ij}$, and the corrected correlation from the synchronous process $c$. This is possible because our correction is performed after obtaining the measured correlation. Our correction is then easier to compute as we do not need to search for individual overlapping intervals to compensate.

\subsection{Correcting for asynchrony}\label{subsec:correction}

Here the Epps effect arises from sampling. For this reason the effect can be corrected. We proceed to discuss a few methods to do so. A key assumption underlying the correction is that there exists an underlying ``true'' and synchronous process which has some correlation quantity we are interested in. The observed samples are then merely realisations from this underlying process.

To measure the correlation $\tilde{\rho}_{\Delta t}^{ij}$ from the observables of the asynchronous process, we use previous tick interpolation coupled with the Realised Covariance (RC) estimator:
$$
\tilde{\Sigma}^{ij}_{T, \Delta t} = \sum_{h=0}^{\lfloor T/\Delta t \rfloor} \left(\tilde{X}_{h \Delta t}^i - \tilde{X}_{(h-1) \Delta t}^i\right) \left(\tilde{X}_{h \Delta t}^j - \tilde{X}_{(h-1) \Delta t}^j\right).
$$
The measured correlation is then
\begin{equation} \label{epps:eq:measured}
    \tilde{\rho}_{\Delta t}^{ij} = \frac{\tilde{\Sigma}^{ij}_{T, \Delta t}}{\sqrt{\tilde{\Sigma}^{ii}_{T, \Delta t} \cdot \tilde{\Sigma}^{jj}_{T, \Delta t}}}.
\end{equation}

\subsubsection*{Arrival time (overlap) correction:}\label{subsubsec:arrivalcorrection}

The first correction method is to correct for the Epps effect directly from the characterisation of the synchronised process Eq. \eqref{epps:eq:11}. To recover the correlation of the true synchronous process we compute:
\begin{equation}\label{epps:eq:13}
    {\rho}_{\Delta t}^{ij} = \tilde{\rho}_{\Delta t}^{ij} \cdot \frac{\sqrt{{\hat{\kappa}_{\Delta t}^{ii} \cdot \hat{\kappa}_{\Delta t}^{jj}}}}{\hat{\kappa}_{\Delta t}^{ij}}.
\end{equation}
This defines the {\it overlap correction}. Here $\hat{\kappa}_{\Delta t}^{ij}$ can be directly estimated using $U^i$ and $U^j$ with discretisation size $\Delta t$. 

\subsubsection*{Flat trade correction:}\label{subsubsec:prevtickcorrection}

The second correction for the Epps effect is given by \cite{BLPP2019}. They take a different approach and assume that the process is observed at $n = \lfloor \frac{T}{\Delta t} \rfloor + 1$ non-random times equispaced over $[0,T]$, {\it i.e.} $0<t_{0,n} < t_{1,n} < ... <t_{n,n} = T$ where $\Delta t = t_{j,n} - t_{j-1,n}$ for $j \geq 1$. They then characterise the observables as
\begin{equation}\label{epps:eq:BLPP}
    \tilde{X}^i_{t_{j,n}} = X^i_{t_{j,n}} \left(1 - B^i_{j,n} \right) + \tilde{X}^i_{t_{j-1,n}} B^i_{j,n},
\end{equation}
where $B^i_{j,n}$ are pairwise-independent triangular arrays of i.i.d. Bernoulli variables. The probability of flat trading is then $\mathbb{E}[{B^i_{j,n}}] = p_i$. In this setting they prove that the Realized Covariance estimator is biased (their Theorem 3.1) to then provide a correction
\begin{equation}\label{epps:eq:14}
    {\rho}_{\Delta t}^{ij} = \tilde{\rho}_{\Delta t}^{ij} \cdot \frac{\left( 1 - \hat{p}_{\Delta t,i} \hat{p}_{\Delta t,j} \right)}{\left( 1 - \hat{p}_{\Delta t,i} \right)\left( 1 - \hat{p}_{\Delta t,j} \right)}.
\end{equation}
Here $\hat{p}_{\Delta t,i}$ is the estimate of the probability of flat trading, $p_i$, given by
\begin{equation}\label{epps:eq:15}
    \hat{p}_{\Delta t,i} = \frac{1}{\lfloor T/\Delta t \rfloor} \sum_{j=1}^{\lfloor T/\Delta t \rfloor} \mathbbm{1}_{\left\{ \tilde{X}_{j \Delta t}^i - \tilde{X}_{(j-1) \Delta t}^i \right\} = 0},
\end{equation}
where $\mathbbm{1}$ is an indicator function. Convergence $\hat{p}_{\Delta t,i} \overset{p}{\longrightarrow} p_i$ is then shown to hold.

We highlight that the assumptions used by \citet{BLPP2019} are not always well suited for illiquid tick-by-tick trade data coupled with previous tick interpolation to achieve Eq. \eqref{epps:eq:BLPP}. Concretely, Assumption 3 of their paper states that the maximum number of consecutive repeated trades is not large relative to the number of data points. This is because the price will only move to a new level when there is a new observable, and then all the grid points between observables will be flat trades. It is therefore difficult to satisfy Assumption 3 using small grid points in illiquid markets with finite data.

\subsubsection*{Hayashi--Yoshida (baseline) correction:}\label{subsubsec:HY}
Here the estimator corrects the effect by accounting for multiple contributions \cite{HY2005} and is defined by
\begin{equation} \label{epps:eq:16}
\begin{aligned}
  \hat{\Sigma}^{ij}_{T}
  &= \sum_{\ell=1}^{\# U^i} \sum_{k=1}^{\# U^j} \left( X_{t_{\ell}^i}^i - X_{t_{\ell-1}^i}^i \right) \left( X_{t_{k}^j}^j - X_{t_{k-1}^j}^j \right) \mathbbm{1}_{\left\{\left(t_{\ell-1}^i, t_{\ell}^i\right] \cap \left(t_{k-1}^j, t_{k}^j\right] \neq \emptyset \right\}},
\end{aligned}
\end{equation}
where $\# U^i$ denotes the cardinality of the set $U^i$. The correlation is then given as
\begin{equation} \label{epps:eq:17}
    {\rho}_{}^{ij} = \frac{\hat{\Sigma}^{ij}_{T}}{\sqrt{\hat{\Sigma}^{ii}_{T} \cdot \hat{\Sigma}^{jj}_{T}}}.
\end{equation}
The Hayashi--Yoshida estimator is independent of the sampling interval $\Delta t$ induced by the estimator from discretising the process into a synchronised process (performed using previous tick interpolation for the RC estimator). This is because Eq. \eqref{epps:eq:16} uses the synchronous process $X^i_t$ where the times correspond to asynchronous transaction times given by the set of arrivals $U^i$. There is no re-sampling or discretisation that induces a time scale $\Delta t$. This means that we cannot use the estimator to investigate the correlations at a specific time scale $\Delta t$. The estimator is independent of $\Delta t$ induced by the estimator. For this reason we will use it as a {\it baseline correction} for comparison. 

Note that the estimator is not independent of the sampling frequencies because it depends on the frequency of arrivals $U^i$. This means that the estimator can be used to investigate different time scales induced by the observables using, for example, $k$-skip sampling. However, it then becomes unclear what time scale ($\Delta t$) the estimator is recovering. Section~\ref{sec:key} will discuss the two time scales in more detail.

Although we do not deal with lead-lag, we highlight that the Hayashi--Yoshida estimator cannot account for lead-lag as it assumes that the correlation between two assets do not extend beyond the interval with full or partial overlap \cite{GO2011}.

\section{Simulation Experiments}\label{sec:experiments}

The experiments focus on comparing the correction of the Epps effect arising from asynchrony from two types of sampling. The first from a homogeneous Poisson process -- \textit{Poisson sampling}. The second from a 2-dimensional Hawkes process -- \textit{Hawkes sampling}. The sampling methods are performed on three types of price models. The first is a standard Brownian price model, the second is a stochastic volatility model contaminated with additive noise and the third is the Hawkes price model \cite{BDHM2013a}.

The experiments are conducted using simulated paths where $T$ is 20 hours (72,000 seconds). The correlations will be investigated at small intervals where $\Delta t$ is measured in seconds. The experiments use a single realisation of the price paths.\footnote{Simulating many paths does not change our results but could enhance the error ribbons.} This is for two reasons. First, this emulates the circumstance with real data where we only have a single realisation. Second, this allows us to focus only on the effect of asynchrony when constructing the error ribbons and ensures that we do not conflate the effect of asynchrony with the sample specific variations from different price paths. 

Each price path is then re-sampled using either a Poisson or Hawkes sampling scheme for 100 replications. The figures plot the mean estimate at each $\Delta t$ over 100 replications for the various correlation and correction estimates. The error ribbons contain 95\% of the estimates at each $\Delta t$ from the replications computed using the Student $t$-distribution with 99 degrees of freedom and the standard deviation of the estimates between the replications at each $\Delta t$.

Our simulations have two types of sampling. First is the sampling of the synchronous process using Poisson or Hawkes sampling to obtain the observables. Second is the sampling from the observables using previous tick interpolation at a particular $\Delta t$ to create a synchronised process which in turn can be used to estimate the measured correlation using Eq. \eqref{epps:eq:measured}, the overlap correction using Eq. \eqref{epps:eq:13}, and the flat trade correction using Eq. \eqref{epps:eq:14}. The two types of sampling relate to the {\it two time scales} induced by the observables and estimators respectively; this is discussed in greater detail in Section~\ref{sec:key}.

\subsection{Brownian price model}\label{subsec:BPM}

\begin{figure*}[ht!]
    \centering
    \subfloat[][Brownian price model with Poisson sampling. ]{{\includegraphics[width=0.5\textwidth]{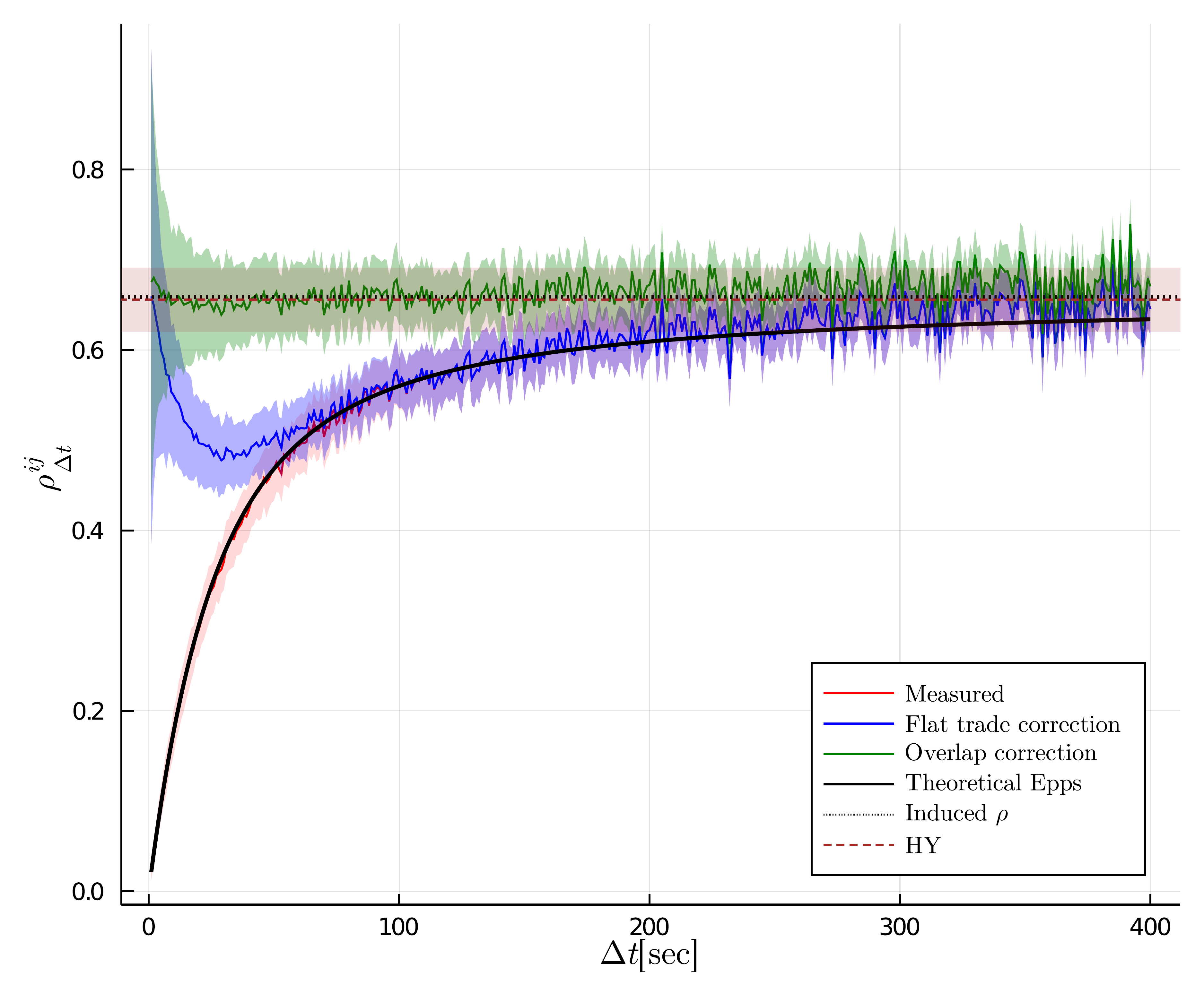}}\label{fig:BPM:a}}
    \subfloat[][Brownian price model with Hawkes sampling. ]{{\label{fig:BPM:b}\includegraphics[width=0.5\textwidth]{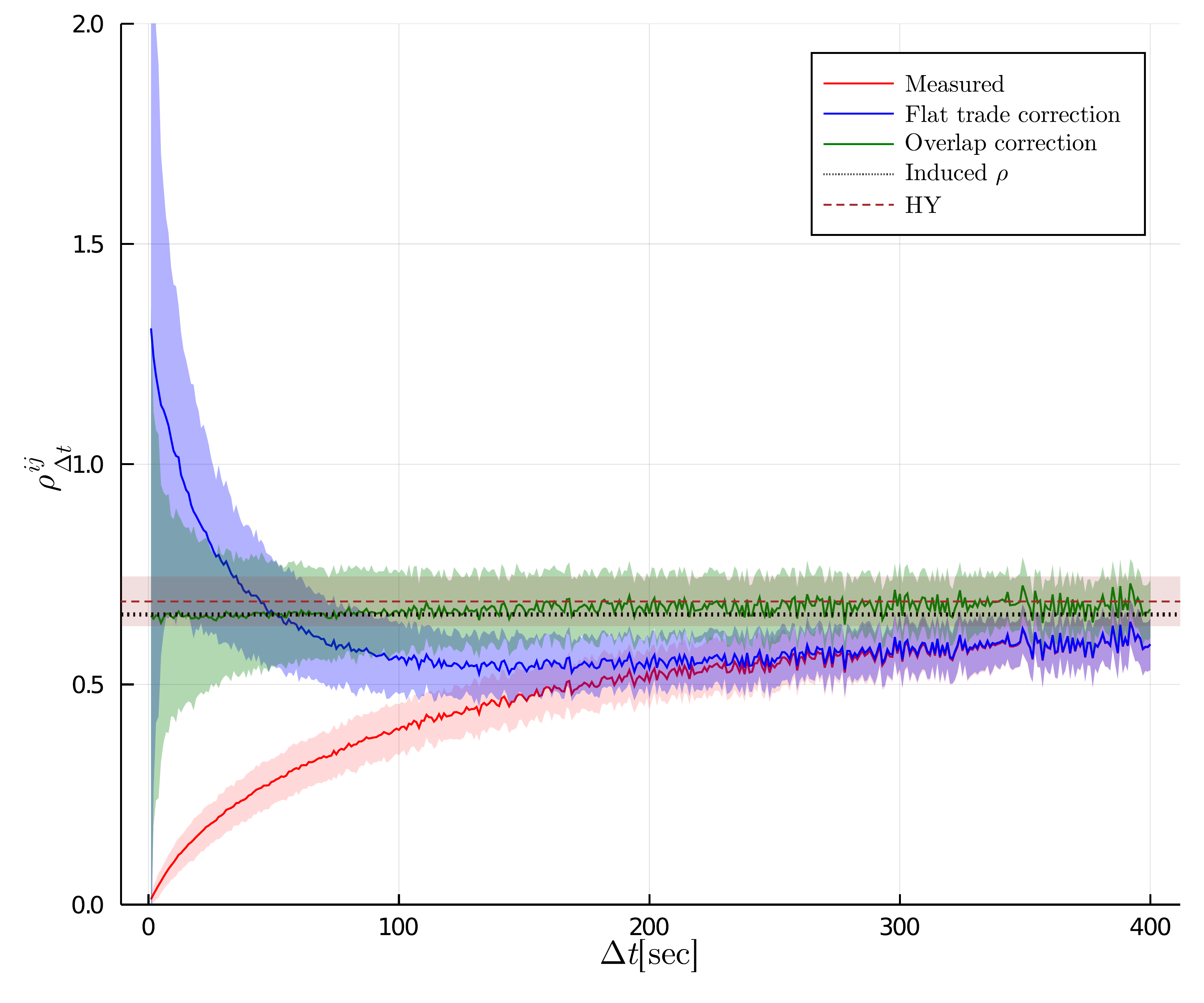}}}
    \caption{A Brownian price model with (a) Poisson sampling and (b) Hawkes sampling is presented: the red line is the measured correlation from the synchronised process calculated using Eq. \eqref{epps:eq:measured}, the blue line is the flat trade correction calculated using Eq. \eqref{epps:eq:14}, and the green line is the overlap correction calculated using Eq. \eqref{epps:eq:13}. The black line is the plot of Eq. \eqref{epps:eq:12}. The horizontal dotted line is the induced correlation of the synchronous system with $\rho \approx 0.65$. Lastly, the horizontal dashed line is the Hayashi--Yoshida estimate, Eq. \eqref{epps:eq:17}.}
\label{fig:BPM}
\end{figure*}

Consider a standard Brownian log-price model satisfying the following SDEs:
$$
\begin{aligned}
    &dX_t^1 = \left(\mu_1 - \frac{\sigma^2_1}{2}\right)dt + \sigma_1dW_t^1, \\
    &dX_t^2 = \left(\mu_2 - \frac{\sigma^2_2}{2}\right)dt + \sigma_2dW_t^2. 
\end{aligned}
$$
Here the infinitesimal correlation between $dW_t^1$ and $dW_t^2$ is $\rho^{12}$, and is set to be approximately $0.65$.\footnote{The value is chosen to compare against the limiting correlation of the Hawkes price model.} The process is simulated using the Euler–Maruyama scheme \footnote{The scheme will be exact for this model, but not for state-dependent drifts and diffusions.} with the discretisation size corresponding to one second intervals. The parameters used are $\mu_1 = \mu_2 = 0.01$, $\sigma^2_1 = 0.1$ and $\sigma^2_2 = 0.2$ (given in daily intervals). The first 500 seconds of this process is shown in Figure~\ref{fig:GBMPricePaths}. 

To obtain the observables, this process is then sampled using a Poisson process with a mean inter-arrival of $1/\lambda = 15$ seconds and a 2-dimensional Hawkes process with $\lambda^1_0 = \lambda^2_0 = 0.015$ and $\boldsymbol{\Phi}$ taking the form
\begin{equation}\label{exp:eq:2}
    \boldsymbol{\Phi} = 
\begin{pmatrix}
0 & \phi^{(s)}    \\
\phi^{(s)} & 0 
\end{pmatrix},
\end{equation}
where $\phi^{(s)} = 0.023 e^{-0.11 t}\mathbbm{1}_{t \in \mathbb{R}^+}$.

Figure~\ref{fig:BPM} compares the correction methods for (a) Poisson and (b) Hawkes sampling on the Brownian price model. The overlap correction and the Hayashi--Yoshida estimator correctly recovers the underlying correlation of the synchronous process. What is interesting is that the flat trade correction does not recover the induced underlying correlation and the correction returns estimates outside the feasible range for correlations under the Hawkes sampling.\footnote{None of the correction methods ensure that corrected correlation estimates lie within $[-1, 1]$.} 

This is curious because the estimator is in fact a consistent estimator as proved by \citet{BLPP2019}. We suspect that the cause of this is due to the long inter-arrivals from our sampling of the observables coupled with a small discretisation size $\Delta t$ from the previous tick interpolation which leads to extended periods of flat trading. These extended periods of flat trading makes the estimates of flat trade correction under Eq. \eqref{epps:eq:15} closer to one. This leads to two possible explanations for why the underlying correlation is not recovered. First, the variance of the flat trade correction spikes when the probabilities of flat trading, $p_1$ and $p_2$, are simultaneously close to one (see Figure 3 of \citet{BLPP2019}). Second, the high probability of flat trading may lead to numerous consecutive flat trades which possibly violates Assumption 3 under which the estimator is constructed \cite{BLPP2019}.

The theoretical Epps effect arising from asynchrony (thick solid line) is plotted for the Poisson sampling since the distribution of $\mathcal{W}_{\Delta t}$ can be recovered in this case. Thus the theoretical Epps effect is given by Eq. \eqref{epps:eq:12}. This theoretical Epps effect is not plotted for the Hawkes sampling as obtaining the distribution of $\mathcal{W}_{\Delta t}$ is not simple in this case. The main feature behind our correction method is that we do not need to obtain the distribution of $\mathcal{W}_{\Delta t}$ since the correction can be directly estimated using Eq. \eqref{epps:eq:10} given $U^i$ and $U^j$.

\subsection{Market microstructure noise model}\label{subsec:JD}

\begin{figure*}[ht!]
    \centering
    \subfloat[][MMN model with Poisson sampling.]{\label{fig:MMN:a}\includegraphics[width=0.5\textwidth]{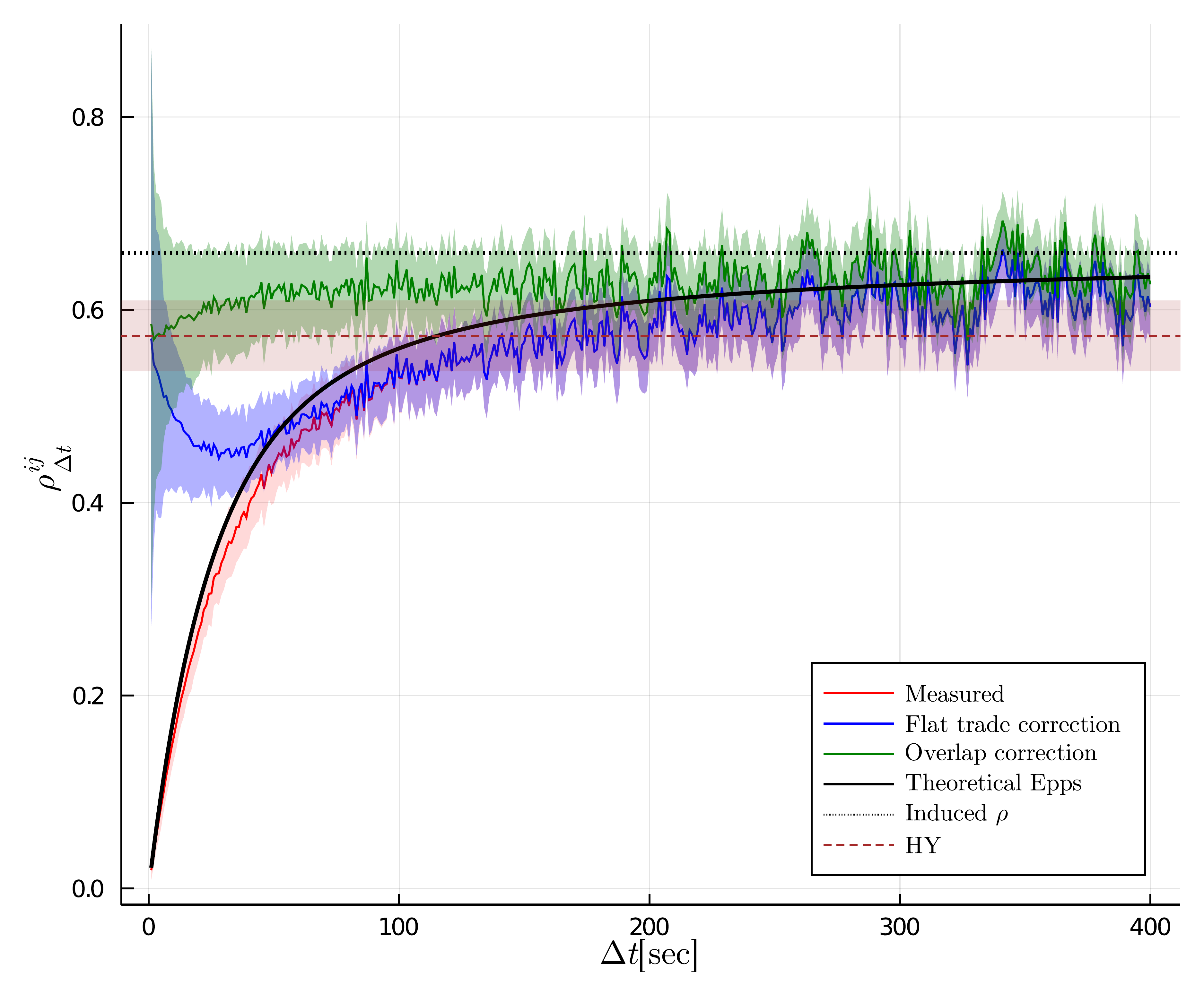}}
    \subfloat[][MMN model with Hawkes sampling.]{\label{fig:MMN:b}\includegraphics[width=0.5\textwidth]{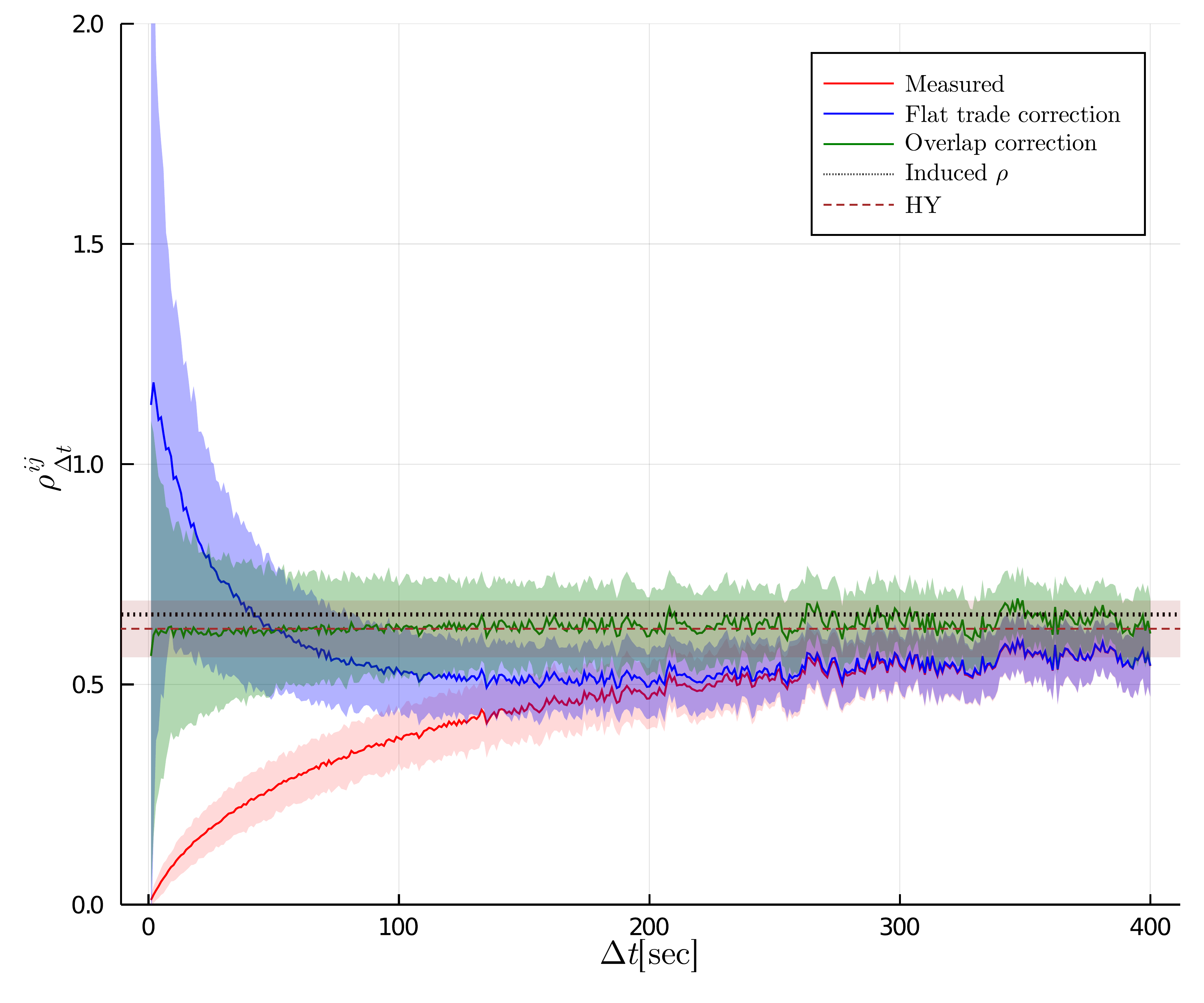}}
    \caption{A market microstructure noise model with (a) Poisson sampling and (b) Hawkes sampling is presented: the red line is the measured correlation from the synchronised process, Eq. \eqref{epps:eq:measured}, the blue line is the flat trade correction, Eq. \eqref{epps:eq:14}, and the green line is the overlap correction, Eq. \eqref{epps:eq:13}. The black line is the plot of Eq. \eqref{epps:eq:12}. The horizontal dotted line is the induced correlation of the synchronous system with $\rho \approx 0.65$. Lastly, the horizontal dashed line is the Hayashi--Yoshida estimate, Eq. \eqref{epps:eq:17}.}
\label{fig:MMN}
\end{figure*}

Consider a more complex {\it ad hoc} model that is still predicated on Brownian motions. We consider a log-price model where the synchronous process $X_t^i$ has been contaminated with additive noise:
$$
\begin{aligned}
    &X_t^1 = P_t^1 + \epsilon_t^1, \quad \epsilon_t^1 \sim \text{NID}\left( 0, h_1^2 \right), \\
    &X_t^2 = P_t^2 + \epsilon_t^2, \quad \epsilon_t^2 \sim \text{NID}\left( 0, h_2^2 \right),
\end{aligned}
$$
where $\epsilon_t^i$ is normal white noise that captures market microstructure effects. Here we assume that the latent process $P_t^i$ satisfies the following SDEs:
$$
\begin{aligned}
    &dP_t^1 = \sigma_{t,1} dW_t^1, \\
    &dP_t^2 = \sigma_{t,2} dW_t^2,
\end{aligned}
$$
where the infinitesimal correlation between $dW_t^1$ and $dW_t^2$ is $\rho^{12}$, and is set to be approximately $0.65$. Here $\sigma_{t,i}$ is a time varying standard deviation that can be decomposed as
$$
\sigma_{t,i} = d_t g_{t,i},
$$
where $d_t$ is a common diurnal pattern given by
$$
d_{t}=C+A e^{-a t}+B e^{-b(1-t)}
$$
and $g_{t,i}$ evolves according to
$$
d g_{t, i}^{2}=k_{i}\left(v_{i}-g_{t, i}^{2}\right) d t+w_{i} g_{t, i} d B_{t}^i.
$$
This model will be henceforth referred to as the market microstructure noise (MMN) model.\footnote{This is the same model used in \citet{BCP2020} but without lead-lag components.} The process is simulated using the Euler–Maruyama scheme with the discretisation size corresponding to one second intervals. The parameters $A=0.75$, $B=0.25$ and $C=0.88929198$ are set in a way such that $1 / T\left(\int_{0}^{T} d_{t}^{2} d t\right)=1$. The remaining parameters are set as $v_1 = 0.01$, $v_2 = 0.02$, $w_1 = w_2 = 0.1$ and $k_1 = k_2 = 10$. The average signal-to-noise ratio between $v_i$ and $h_i^2$ is set to be one-to-one and the leverage effect between $d B_{t}^i$ and $dW_t^i$ is set to be $0.01$.

To obtain the observables, we sample the MMN model with a Poisson process with a mean inter-arrival of $1/\lambda = 15$ seconds and a 2-dimensional Hawkes process taking the form of Eq. \eqref{exp:eq:2} with parameters $(\lambda_0^1=\lambda_0^2, \alpha^{(s)}, \beta) = (0.015, 0.023, 0.11)$.

Figure~\ref{fig:MMN} compares the correction methods for (a) Poisson and (b) Hawkes sampling on the MMN model. We see that in Figure~\ref{fig:MMN:a} that all the estimates are lower compared to the estimates from Figure~\ref{fig:BPM:a}. The lower correlation achieved is a result of the contamination from noise. It is known that the realised volatility estimator overestimates the volatility when the process is contaminated with additive noise \cite{ZMA2005}. This increased volatility estimate leads to a larger normalisation factor and thus the correlation estimates are lower than before.

What is interesting is that the estimates in Figure~\ref{fig:MMN:b} are not as affected by noise when compared to Figure~\ref{fig:MMN:a}. The reason behind this is because the specification of the Hawkes sampling in Eq. \eqref{exp:eq:2} results in longer inter-arrivals compared to the Poisson sampling. This hints at the fact that the time scale induced by the observables play a role in our ability to correct for asynchrony. This result forms the basis for our ability to discriminate between the processes which we will discuss in Section~\ref{sec:key}.

\subsection{Hawkes price model}\label{subsec:HPM}

\begin{figure}[htp]
    \subfloat[The Epps effect from a Hawkes price model.]{\label{fig:HawkesSynEpps(a)}\includegraphics[width=0.49\textwidth]{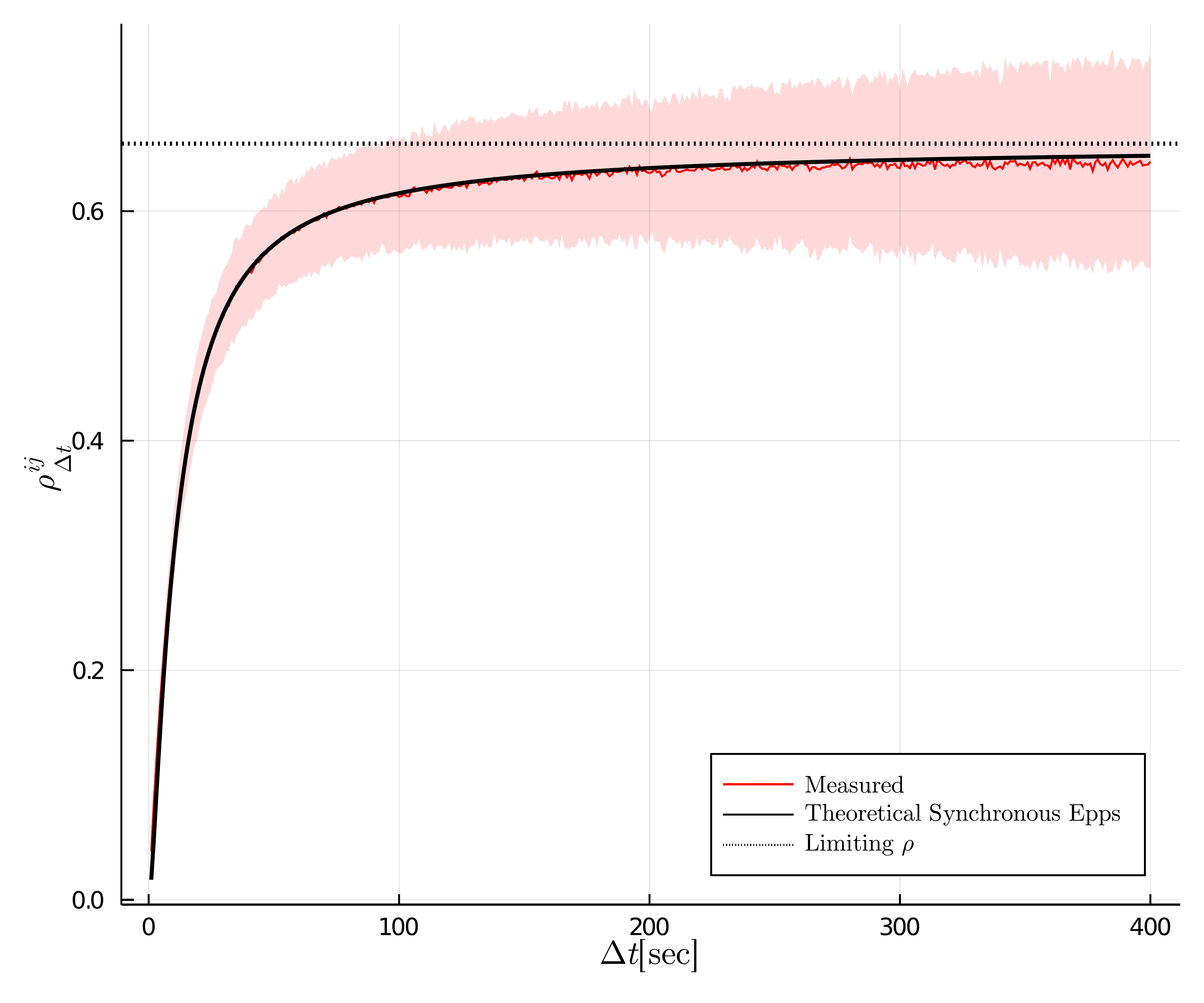}}\\
    \subfloat[Full plot of 20 hours with 1 hour inset.]{\stackinset{l}{0.11\textwidth}{t}{0.205\textwidth}{\label{fig:HPP:b}\includegraphics[width=0.24\textwidth]{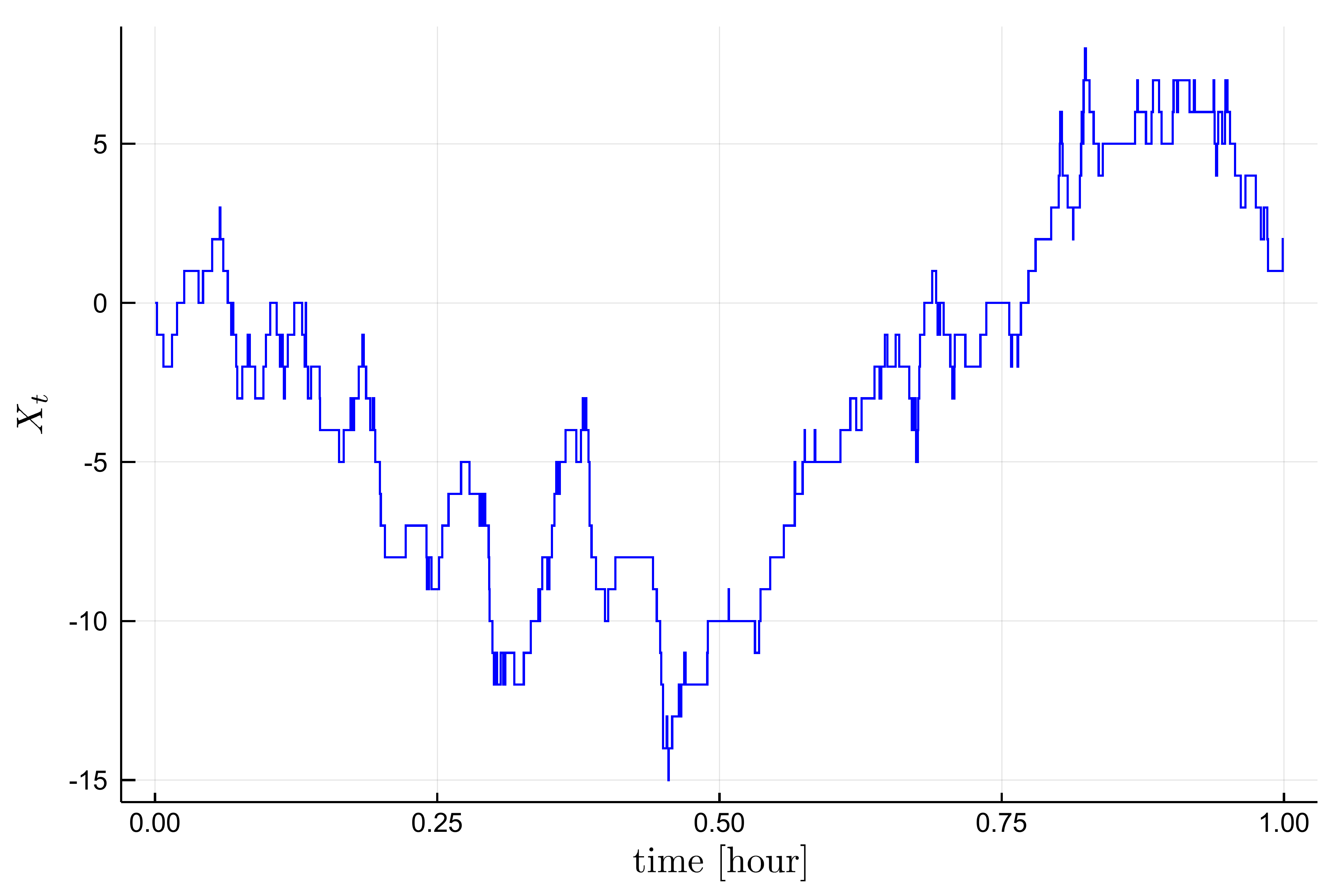}}{\label{fig:HPP:a}\includegraphics[width=0.49\textwidth]{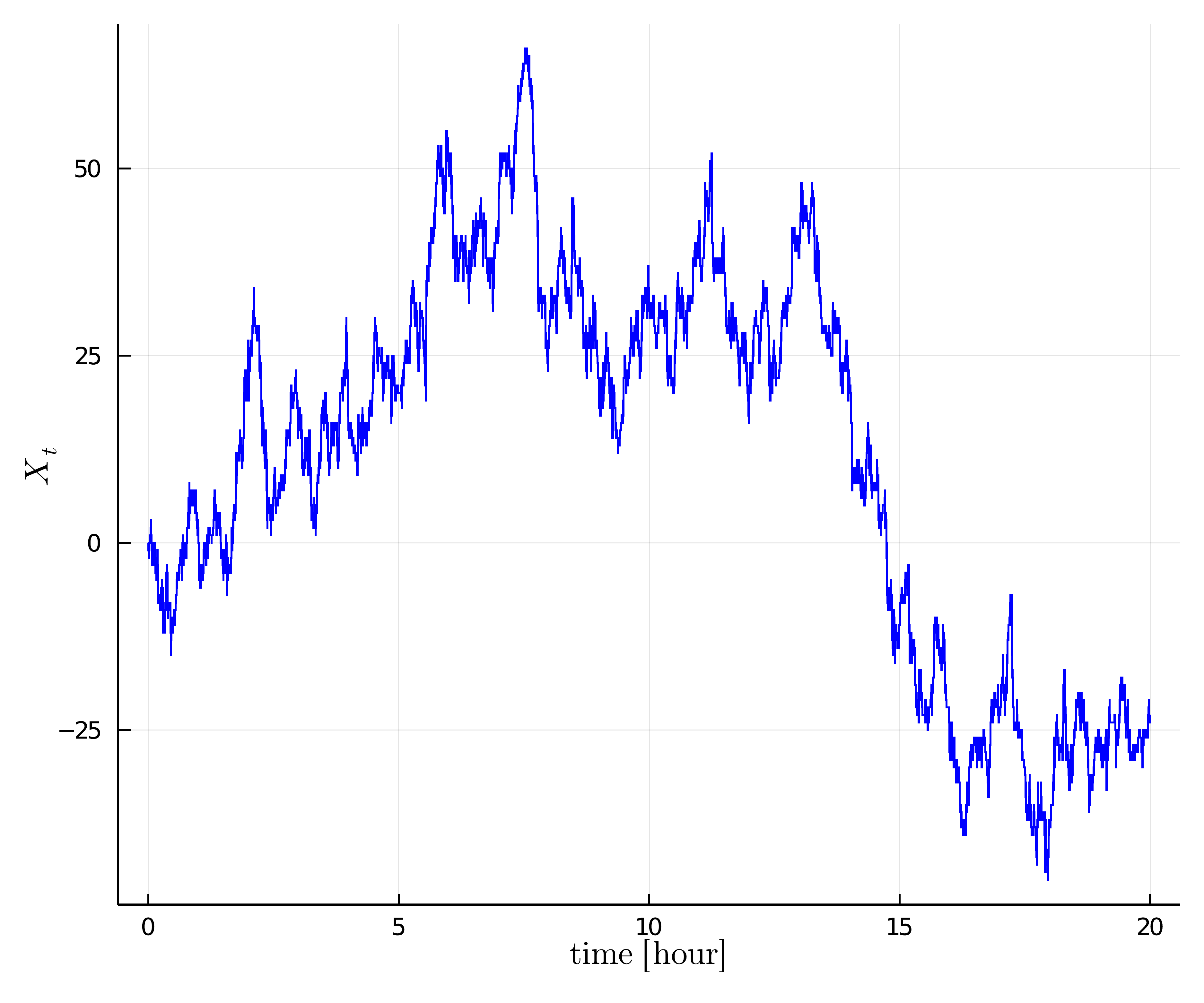}}}
    \caption{The Epps effect is demonstrated from the Hawkes price model on the left, where the red line is the measured correlation from Eq. \eqref{epps:eq:measured} from \emph{synchronous} samples. The black line is the plot of Eq. \eqref{exp:eq:5} and the dotted horizontal line is the correlation as $\Delta t \rightarrow \infty$. Using the parameters $(\mu, \alpha^{(r)}, \alpha^{(c)}, \beta) = (0.015, 0.023, 0.05, 0.11)$ we have that $\rho_{\Delta t}^{12} \approx 0.65$ as $\Delta t \rightarrow \infty$. The Hawkes price model is plotted on the right for the full 20 hours where the path looks like that from a diffusion process, and an inset that zooms in the first hour shows jumps. }
    \label{fig:HawkesSynEpps}
\label{fig:HPP}
\end{figure}

Consider the price model introduced by \cite{BDHM2013a} constructed using interconnected events from a Hawkes process. Let the bivariate log-price be
$$
\begin{aligned}
    &X_t^1 = X_0^1 + N_1(t) - N_2(t), \\
    &X_t^2 = X_0^2 + N_3(t) - N_4(t),
\end{aligned}
$$
where $\{N_m(t)\}_{m=1}^4$ is a 4-dimensional mutually exciting Hawkes process with the kernel $\boldsymbol{\Phi}$ taking the form:
\begin{equation}\label{exp:eq:4}
    \boldsymbol{\Phi} = 
\begin{pmatrix}
0 & \phi^{(r)} & \phi^{(c)} & 0   \\
\phi^{(r)} & 0 & 0 & \phi^{(c)}   \\
\phi^{(c)} & 0 & 0 & \phi^{(r)}   \\
0 & \phi^{(c)} & \phi^{(r)} & 0
\end{pmatrix}.
\end{equation}
The interpretation of kernel components in Eq. \eqref{exp:eq:4} is as follows: $\phi^{(r)}$ imitates mean reversion because an uptick in $X^1$ by $N_1$ will lead to an increased intensity in the down tick $N_2$---allowing the price level to revert (similarly for $X^2$ through $N_3$ and $N_4$). On the other hand $\phi^{(c)}$ induces a correlation between the prices by connecting the two prices since an uptick in $X^1$ by $N_1$ will lead to an increased intensity in the uptick of $X^2$ through $N_3$ (similarly for down ticks through $N_2$ and $N_4$).

We use the same specification and parameters as \cite{BDHM2013a} for the construction of kernel in Eq. \eqref{exp:eq:4} with $\lambda_0^m = \mu, \forall m$, $\phi^{(r)} = \alpha^{(r)} e^{-\beta t}\mathbbm{1}_{t \in \mathbb{R}^+}$ and $\phi^{(c)} = \alpha^{(c)} e^{-\beta t}\mathbbm{1}_{t \in \mathbb{R}^+}$. Therefore the parameters are given as $(\mu, \alpha^{(r)}, \alpha^{(c)}, \beta) = (0.015, 0.023, 0.05, 0.11)$.

Figure~\ref{fig:HPP:a} plots a single realisation for the price model. Figure~\ref{fig:HPP:a} looks like a diffusive model similar to that found in Figure~\ref{fig:GBMPricePaths}. However, when we zoom into the path (as shown in the inset) we can see the jumps, and that the price model is composed of discrete events. Over long horizons these events start behaving like diffusion processes. Limit theorems for Hawkes price models have been derived \cite{BDHM2013b}.

\citet{BDHM2013a} formulate the Epps effect using the Hawkes price model where the covariance matrix is a function of $\Delta t$
$$
\begin{aligned}
\frac{C_{\Delta t}^{11}}{\Delta t}&= \Lambda+\frac{R C_{1}}{2 G_{1}}+\frac{R C_{2}}{2 G_{2}} + R \frac{C_{2} G_{1}^{2} (1- \mathrm{e}^{-\Delta t G_{2}})
+ G_{2}^{2} (Q_{1}  \mathrm{e}^{-\Delta t G_{1}} - C_{1}) }
{2 G_{2}^{2} G_{1}^{2} \Delta t}, \nonumber \\
\frac{C_{\Delta t}^{12}}{\Delta t}&= \frac{R C_{2}}{2 G_{2}} -\frac{R C_{1}}{2 G_{1}}+R\frac{
 C_{1} G_{2}^{2}(1- \mathrm{e}^{-G_{1} \Delta t})
+C_{2} G_{1}^{2} ( \mathrm{e}^{-G_{2} \Delta t}-1)}{2 G_{2}^{2} G_{1}^{2} \Delta t}. \nonumber
\end{aligned}
$$
Here the parameters are:
$$
\begin{aligned}
\Lambda &= \frac{\mu}{1 - \Gamma_{12} - \Gamma_{13}},   \\
R &=\frac{\beta \mu}{\Gamma_{12}+\Gamma_{13}-1},    \\
C_{1} &=\frac{\left(2+\Gamma_{12}+\Gamma_{13}\right)\left(\Gamma_{12}+\Gamma_{13}\right)}{1+\Gamma_{12}+\Gamma_{13}},  \\
C_{2} &=\frac{\left(2+\Gamma_{12}-\Gamma_{13}\right)\left(\Gamma_{12}-\Gamma_{13}\right)}{1+\Gamma_{12}-\Gamma_{13}},  \\
Q_{1}=Q_{4}&=\frac{-\mu\left(\Gamma_{12}^{2}+\Gamma_{12}-\Gamma_{13}^{2}\right)}{\left(\left(\Gamma_{12}+1\right)^{2}-\Gamma_{13}^{2}\right)\left(1-\Gamma_{12}-\Gamma_{13}\right)},   \\
Q_{2}=Q_{3}&=\frac{-\mu \Gamma_{13}}{\left(\left(\Gamma_{12}+1\right)^{2}-\Gamma_{13}^{2}\right)\left(1-\Gamma_{12}-\Gamma_{13}\right)},
\end{aligned}
$$
and 
\begin{equation}
G_{1}=\beta\left(1+\Gamma_{12}+\Gamma_{13}\right), \quad G_{2}=\beta\left(1+\Gamma_{12}-\Gamma_{13}\right). \nonumber
\end{equation}
The underlying correlation for the synchronous Hawkes price model is then
\begin{equation}\label{exp:eq:5}
    \rho_{\Delta t}^{12} = \frac{C_{\Delta t}^{12}}{C_{\Delta t}^{11}}.
\end{equation}
In the limit we have \cite{BDHM2013a}
$$
\lim_{\Delta t \to \infty} \rho_{\Delta t}^{12} = \frac{2 \Gamma_{13}\left(1+\Gamma_{12}\right)}{1+\Gamma_{13}^{2}+2 \Gamma_{12}+\Gamma_{12}^{2}}.
$$

Figure~\ref{fig:HawkesSynEpps} demonstrates the Epps effect arising from the Hawkes price model. A crucial feature here is that the Epps effect arises from \emph{synchronous} samples. There are no statistical causes such as asynchrony or tick-size nor is the effect of lead-lag contributing towards the Epps effect. Under this model it is more accurate to think of the Epps effect as the emergence of correlation as $\Delta t$ increases rather than a decay in correlation. This is a crucial insight despite the conceptual simplicity of the model. 

This is because at small time intervals the events are merely random. The connection between the events can only be detected when there are sufficient events in a given interval. Meaning that a system created by events takes a finite time to correlate and requires the concept of intervals to be externally imposed by an external observer to give the notion of correlation meaning. This is the fundamental difference between this model and those considered earlier. Correlation from the Hawkes price model is a result of the prices being coupled between discrete and connected events, whereas the models predicated on Brownian motions have an underlying correlation at infinitesimal scales written in {\it a-priori} by the system observers to then later be estimated and fitted to observed re-sampled data.

\begin{figure*}[htp]
    \centering
    \subfloat[Hawkes price model with Poisson sampling.]{\label{fig:HPM:a}\includegraphics[width=0.5\textwidth]{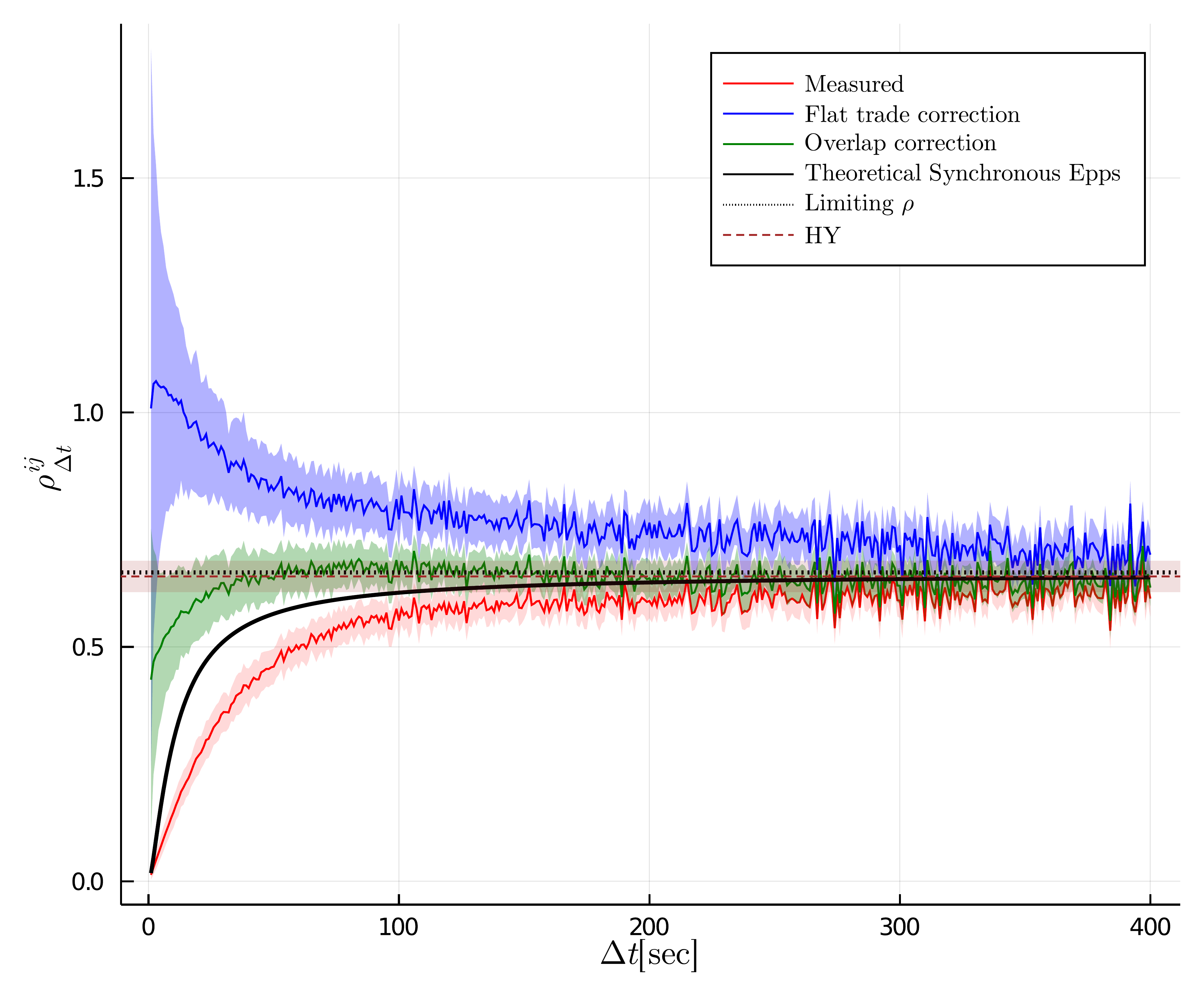}}~~
    \subfloat[Hawkes price model with Hawkes sampling.]{\label{fig:HPM:b}\includegraphics[width=0.5\textwidth]{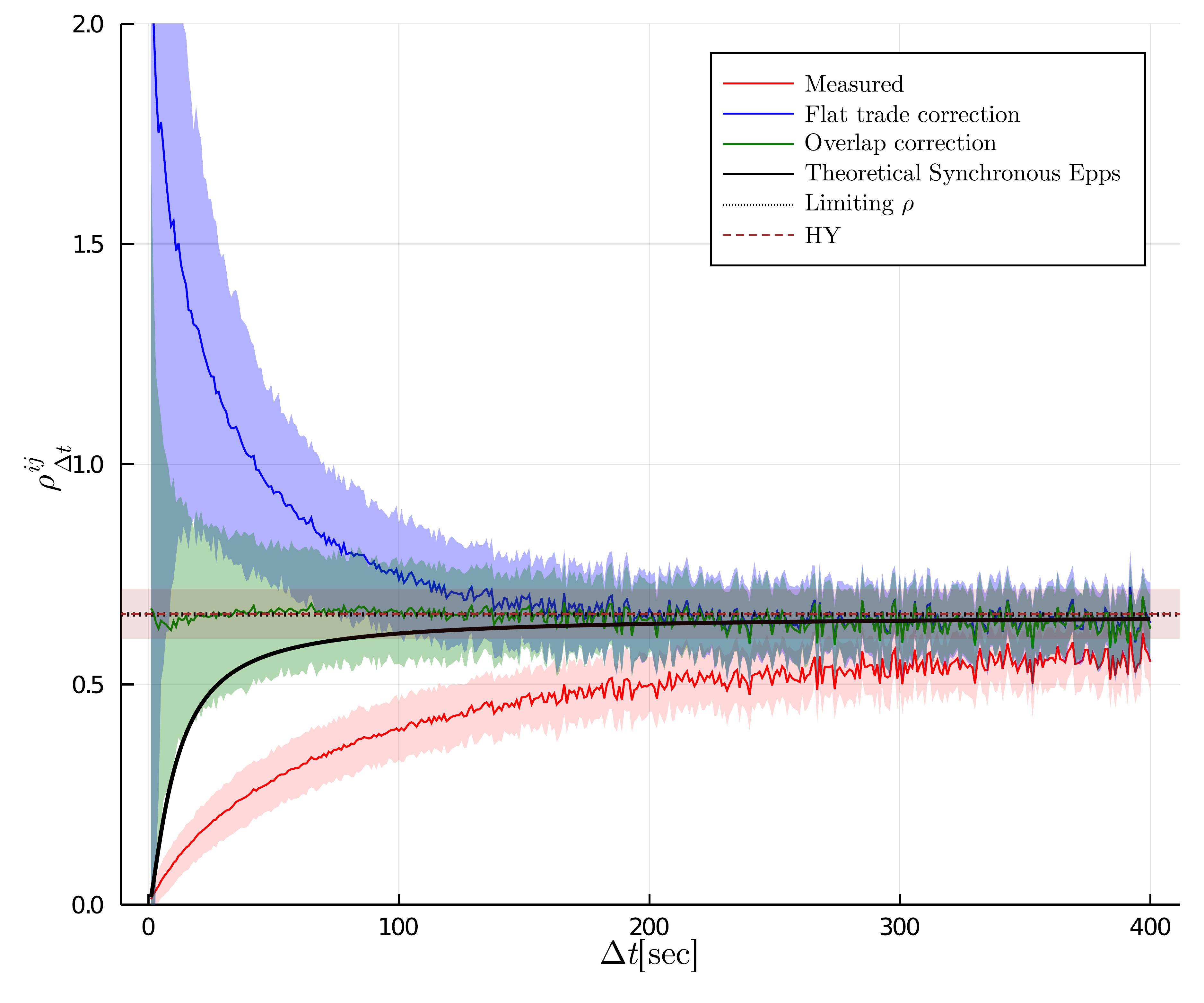}}
    \caption{The Hawkes price model with (a) Poisson sampling and (b) Hawkes sampling is presented: the red line is the measured correlation from the synchronised process, Eq. \eqref{epps:eq:measured}, the blue line is the flat trade correction, Eq. \eqref{epps:eq:14}, and the green line is the overlap correction, Eq. \eqref{epps:eq:13}. The black line is the plot of Eq. \eqref{exp:eq:5}. The horizontal dotted line is the limiting correlation when $\Delta t \rightarrow \infty$ which means $\rho_{\Delta t}^{12} \approx 0.65$. Lastly the horizontal dashed line is the Hayashi--Yoshida estimator, Eq. \eqref{epps:eq:17}.}
\label{fig:HPM}
\end{figure*}

The observables are again obtained by sampling the Hawkes price model using a Poisson process with a mean inter-arrival of $1/\lambda = 15$ seconds and a 2-dimensional Hawkes process taking the form of Eq. \eqref{exp:eq:2} with parameters $(\lambda_0^1=\lambda_0^2, \alpha^{(s)}, \beta) = (0.015, 0.023, 0.11)$.

Figure~\ref{fig:HPM} compares the correction methods for (a) Poisson and (b) Hawkes sampling on the Hawkes price model. We see that the flat trade correction overcorrects the correlation for both cases. This is because the Hawkes price model only jumps to a new price state when an event occurs. Moreover, an uptick is likely followed by a down tick due to the mean reverting property. This interaction can lead to an amplification of an extended period of flat trades which results in the estimate of the probability of flat trading Eq. \eqref{epps:eq:15} being closer to one. 

The measured correlation from the synchronised process exhibits a further Epps effect relative to the theoretical synchronous Epps effect (thick line) which is expected. This difference is naturally attributed towards the asynchronous sampling.

What is interesting is that in both Figures~\ref{fig:HPM:a} and \ref{fig:HPM:b}, the Hayashi--Yoshida estimates recover the limiting correlation but the overlap correction only recovers the limiting correlation in Figure~\ref{fig:HPM:b}. The overlap correction in Figure~\ref{fig:HPM:a} under estimates the limiting correlation and over estimates the synchronous Epps effect. This is curious because if we are correcting for asynchrony then we should be recovering the correlation of the theoretical synchronous Epps effect. This hints again at the fact that the implicit time scales induced by the choices of observables play an important role in our ability to correct for asynchrony.

\section{The residual Epps effect}\label{sec:key}

The key realisation is that there are in fact at least two time scales at play. First is the time scale induced by the observables. In our experimental context this is the sampling of the synchronous process to obtain the observables. Second is the time scale induced by the estimator. This is induced by the previous tick interpolation at a particular discretisation size $\Delta t$ to create a synchronised process. 

Previous tick interpolation is not the only way an estimator can induce a time scale. The Malliavin--Mancino Fourier estimator \cite{MM2002,MM2009} directly uses the observables without the need to create a synchronised process. However, the estimator induces a time scale through the number of Fourier coefficients of the price process it uses to estimate the Fourier coefficients of the volatility process \cite{PCEPTG2020a}. It is this property that was exploited to investigate the Epps effect using the Malliavin--Mancino Fourier estimator \cite{RENO2001,PI2007}.

The interplay between the two time scales and the underlying process will determine our ability to correct for asynchrony. Concretely, the time scale induced by the observables will not affect our ability to correct for asynchrony on the Brownian price model but it will affect our ability to correct for asynchrony on the MMN and Hawkes price model.

\subsection*{Experiment \#1: Hayashi--Yoshida}

\begin{figure*}[htp]
    \centering
    \subfloat[Brownian price model.]{\label{fig:GBMPriceModelHYSamplingFreq:a}\includegraphics[width=0.33\textwidth]{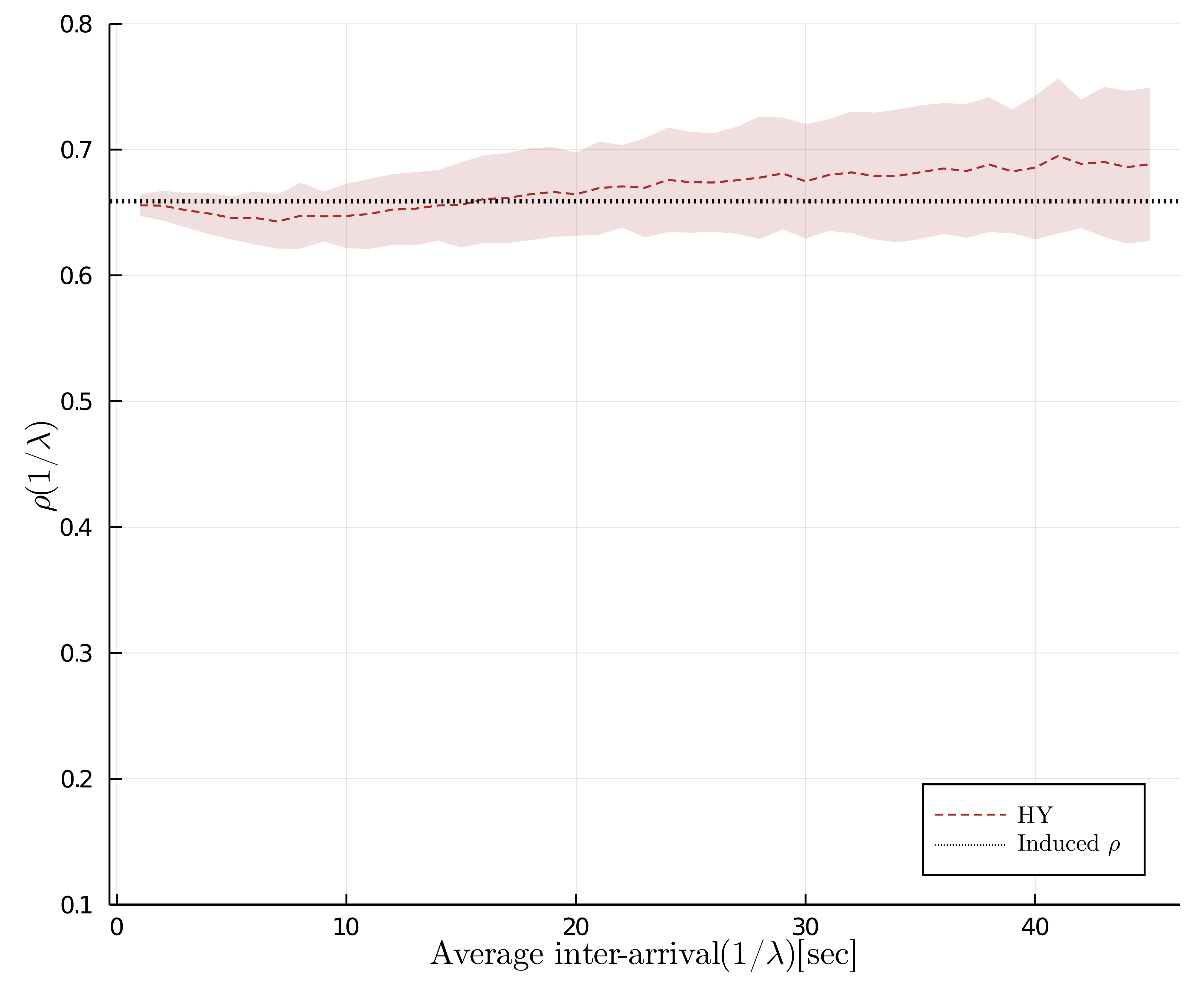}}~~
    \subfloat[MMN model.]{\label{fig:SVPriceModelHYSamplingFreq:b}\includegraphics[width=0.33\textwidth]{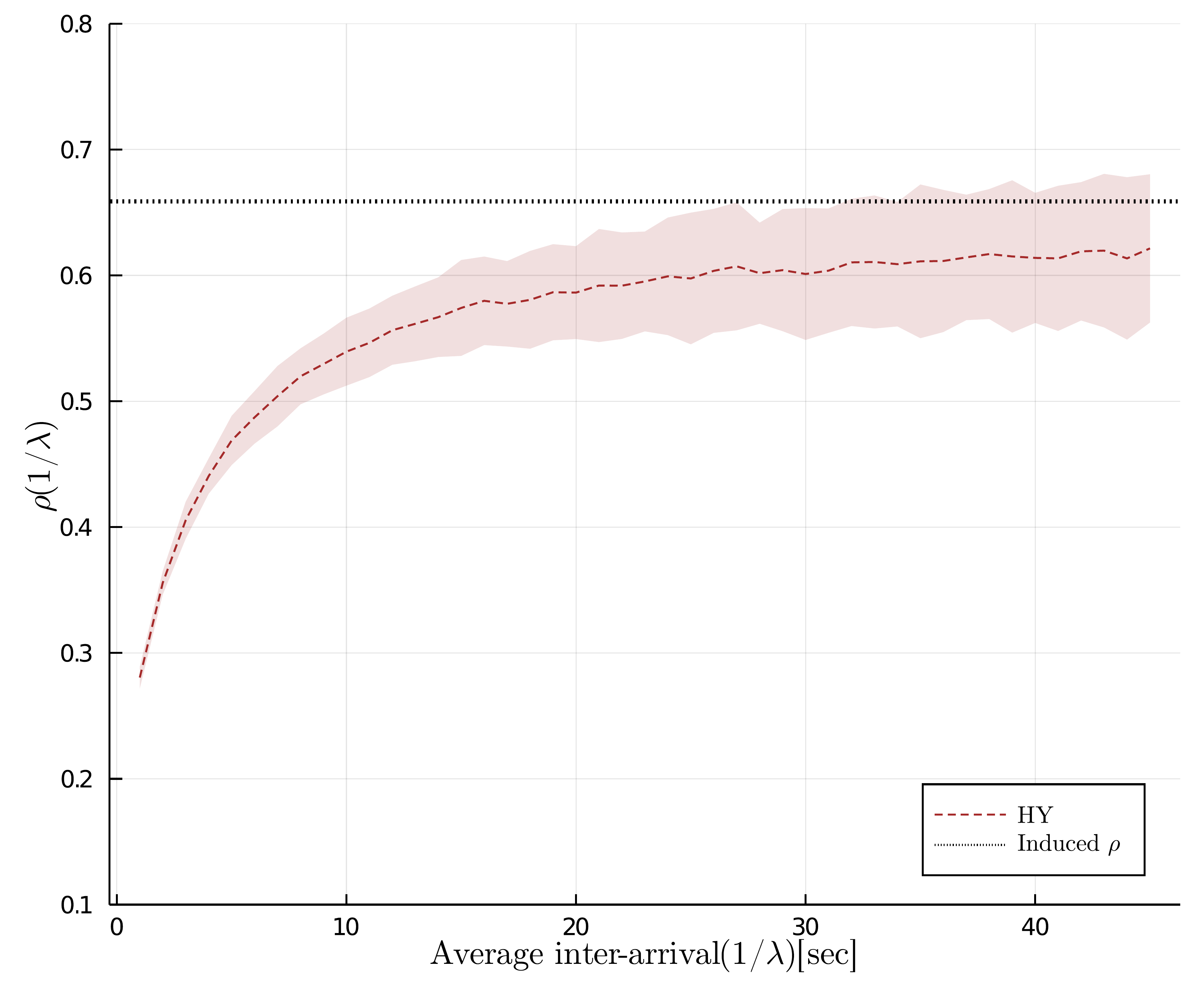}}~~
    \subfloat[Hawkes price model.]{\label{fig:HawkesPriceModelHYSamplingFreq:c}\includegraphics[width=0.33\textwidth]{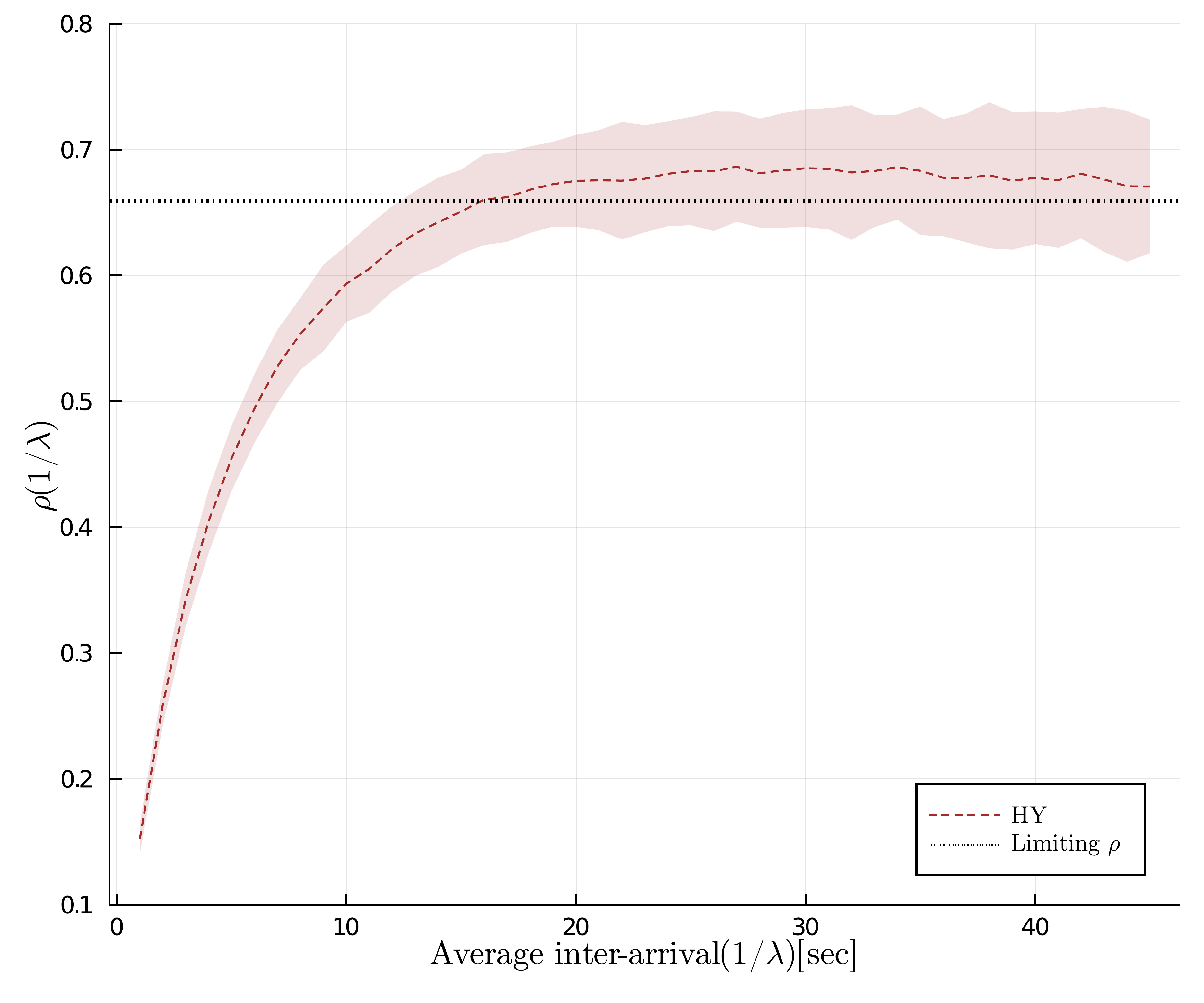}}
    \caption{The Hayashi--Yoshida estimate (dashes lines) given in Eq. \eqref{epps:eq:17} plotted as a function of the mean inter-arrival for (a) the Brownian price model, (b) the MMN model and (c) the Hawkes price model. The limiting/induced $\rho$ (horizontal dotted line) is approximately 0.65 as before. We see how the time scale induced by the observables can affect the correction of asynchrony for the different underlying processes.}
\label{fig:HawkesPriceModelHYSamplingFreq}
\end{figure*}

Let us consider some additional experiments to demonstrate this point. The Hayashi--Yoshida estimator is a unique estimator in that it does not induce a time scale $\Delta t$ when estimating the correlation. It recovers the underlying correlation directly from the observables. This gives us the ability to investigate the effect of the time scale induced by the observables without the interaction from the time scale induced by the estimator.

We simulate a single realisation of the price paths from each of the models investigated in Section~\ref{sec:experiments}. The time scale induced by the observables are controlled using the Poisson sampling with the average inter-arrivals ranging from 1 to 45 seconds. Each price model is then re-sampled with the different average inter-arrivals 100 times. Figure~\ref{fig:HawkesPriceModelHYSamplingFreq} plots the mean Hayashi--Yoshida estimate on (a) the Brownian price model, (b) the MMN model and (c) the Hawkes price model as a function of the average inter-arrival time with error ribbons containing 95\% of the estimates at each $1/\lambda$ over the 100 replications.

We see in Figure~\ref{fig:HawkesPriceModelHYSamplingFreq} that the Hayashi--Yoshida estimator recovers the induced correlation of the Brownian price model regardless of the time scale induced by the observables.\footnote{The discretisation in the simulation of the diffusion process should be finer than the average inter-arrivals to avoid insufficient granularity in the diffusion process.} On the other hand we see that the estimator presents an Epps like effect as a function of the average inter-arrival time for the MMN and Hawkes price model.

However, the Epps like effect from the MMN and Hawkes price models arise because of very different reasons. The decay in correlations from the MMN model is a result of the signal-to-noise ratio at the different time scales induced by the observables. On shorter time scales the log-return observed from transactions is mostly composed of market microstructure noise and brings little information regarding the volatility of the price process, whereas on longer time scales the amount of market microstructure noise remains constant while the informational content of volatility increases \cite{AMZ2005}. This leads to an overestimation of volatility on shorter time scales which results in the decay of correlations. On the other hand the Epps like effect from the Hawkes price model is the result of the observables inducing a different time scale average over the underlying events. This means we recover a correlation estimate from the synchronous process over a mixture of time scales from the observables.

This means we can discriminate the underlying process of an unknown system by re-sampling the process at different average inter-arrivals and estimating the Hayashi--Yoshida estimates.\footnote{Our experiments are not formal hypothesis tests but a heuristic approach aimed at determining which implicit model assumption is better satisfied through the stylised facts.} If the plot is flat for various values of $1/\lambda$ then the underlying process is a diffusion-based process. If the plot presents decaying correlation for smaller $1/\lambda$ then the underlying process is either a diffusion-based process but contaminated with {\it ad hoc} noise or it is composed of discrete connected events (a system where correlation is then emergent).

\subsection*{Experiment \#2: Overlap correction}

\begin{figure*}[htp]
    \centering
    \subfloat[MMN model.]{\label{fig:SVPriceModelwDiffSamplingFreq:a}\includegraphics[width=0.5\textwidth]{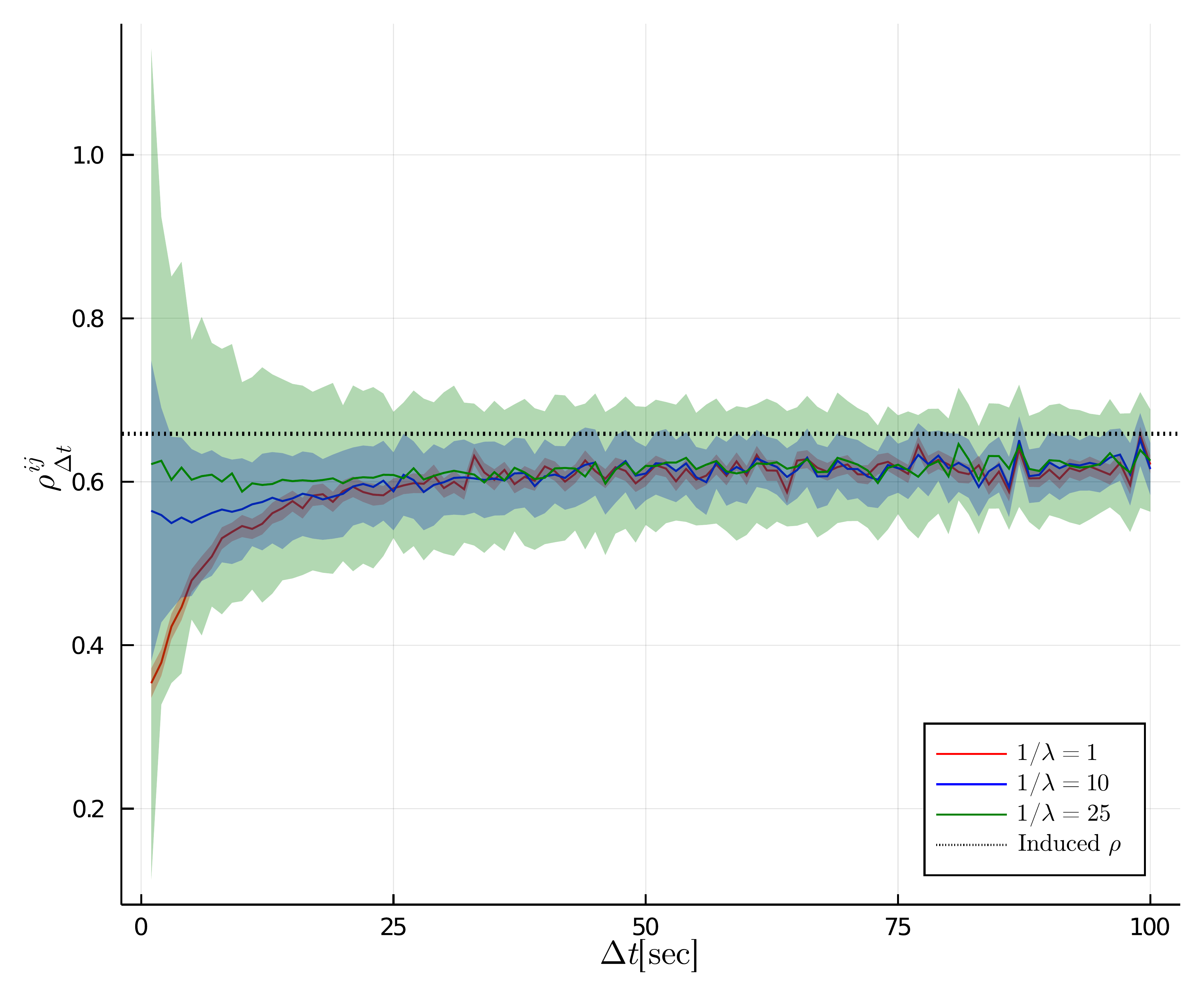}}~~
    \subfloat[Hawkes price model.]{\label{fig:HawkesPriceModelwDiffSamplingFreq:b}\includegraphics[width=0.5\textwidth]{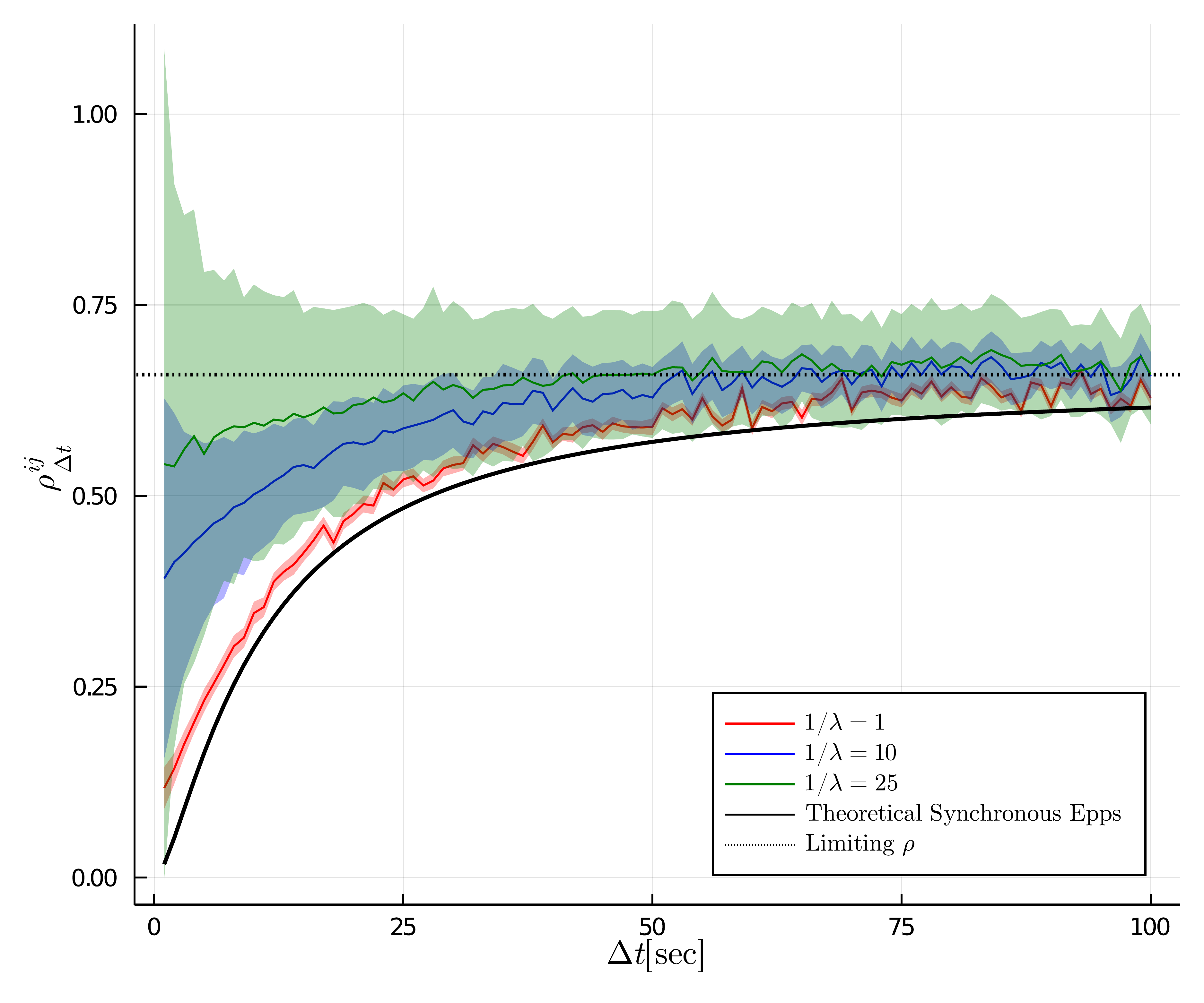}}\caption{The overlap correction Eq. \eqref{epps:eq:13} for three sampling frequencies of observables are plotted. The three sampling frequencies chosen are: $1/\lambda = 1, 10, 25$ which corresponds to the red, blue, green lines respectively. The thick line is the theoretical synchronous Epps effect given by Eq. \eqref{exp:eq:5} and the horizontal dotted line is the induced/limiting $\rho$ set to be approximately 0.65 as before. We see the interplay between the time scale induced by the observables and the time scale induced by the estimator.}
\label{fig:HawkesPriceModelwDiffSamplingFreq}
\end{figure*}

In the next experiment we plot the actual residual Epps curves which allows us to investigate the interplay between the two time scales induced by the observables and the estimator on the MMN and Hawkes price model. Once again the time scale induced by the observables are controlled using the Poisson sampling with average inter-arrivals set as $1/\lambda = 1, 10, 25$. In this case there is no need to perform the experiment with the Brownian price model. This is because with the Poisson sampling, the Epps effect is characterised exactly by Eq. \eqref{epps:eq:12} and therefore the residual Epps effect here will be flat and recover the correct underlying correlation for the various inter-arrivals.

In figure~\ref{fig:HawkesPriceModelwDiffSamplingFreq} we see that in both the MMN and Hawkes price model that the overlap correction tends towards the induced/limiting correlation as $1/\lambda$ increases, whereas the overlap correction presents the residual Epps effect as $1/\lambda$ decreases. 

We see that our ability to recover the induced correlation becomes problematic when $1/\lambda$ becomes smaller for the MMN model. This is because the signal-to-noise ratio is larger on shorter time scales. On the other hand we see that we start to recover the correlation from the synchronous case when $1/\lambda$ is very small for the Hawkes model. This is expected since the asynchrony correction is supposed to recover the correlation from the synchronous process. This confirms that indeed the time scale induced by the observables do in fact play a fundamental role in our ability to correct for asynchrony.

This experiment also gives us a method to discriminate the underlying process of an unknown system. This is achieved by sampling the unknown system with different average inter-arrivals and then plotting the residual Epps curves using the overlap correction. If the residual Epps curves are flat for all values of $1/\lambda$ then the underlying process is a diffusion-based process. If the residual Epps curves decay more for smaller $1/\lambda$ then the underlying process is either a diffusion-based process but contaminated with {\it ad hoc} noise or is composed of discrete connected events.

\subsection*{Experiment \#3: k-skip Hayashi--Yoshida}

\begin{figure*}[h]
    \centering
    \subfloat[Brownian price model.]{\label{fig:k_skip:a}\includegraphics[width=0.33\textwidth]{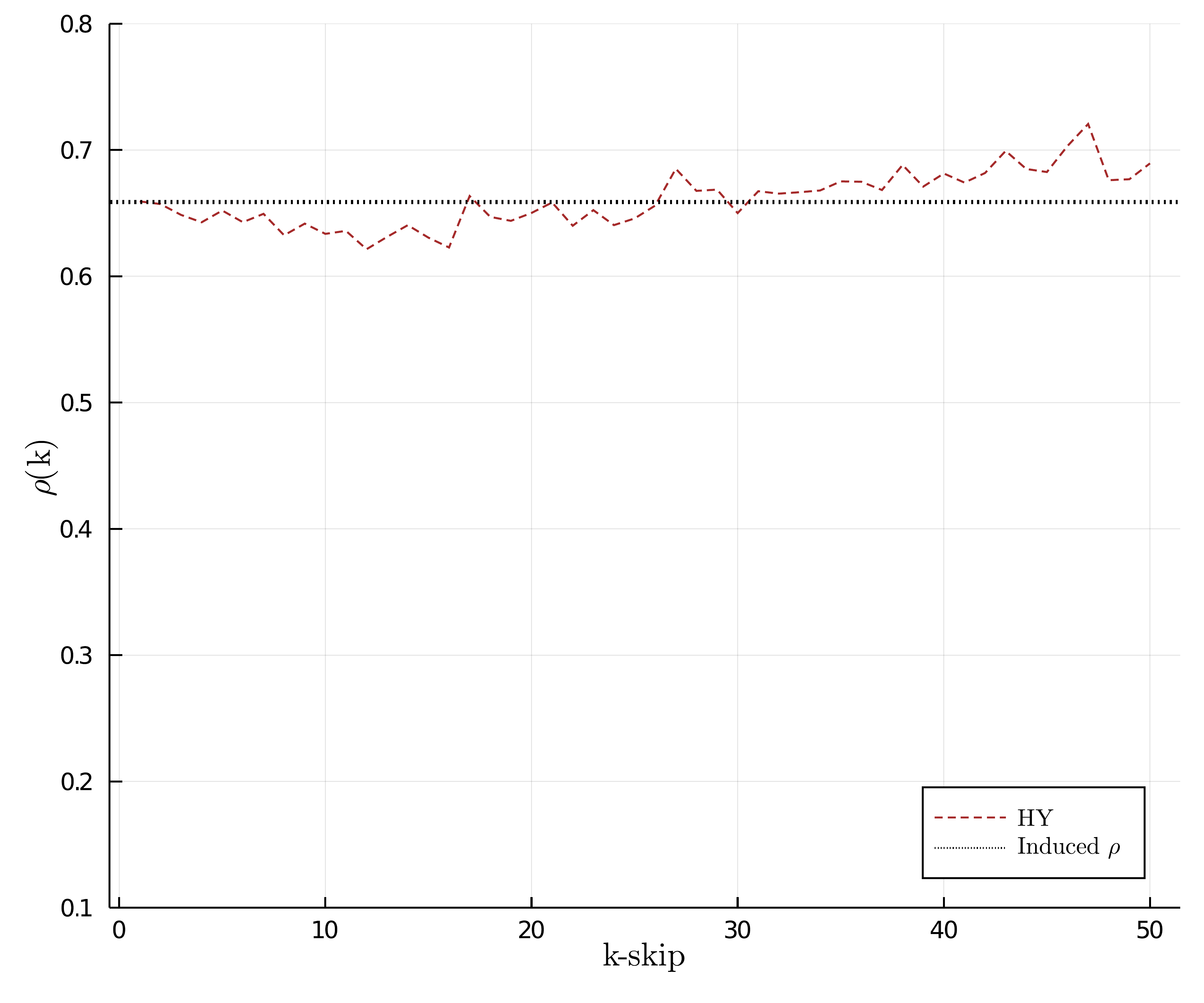}}~~
    \subfloat[MMN model.]{\label{fig:k_skip:b}\includegraphics[width=0.33\textwidth]{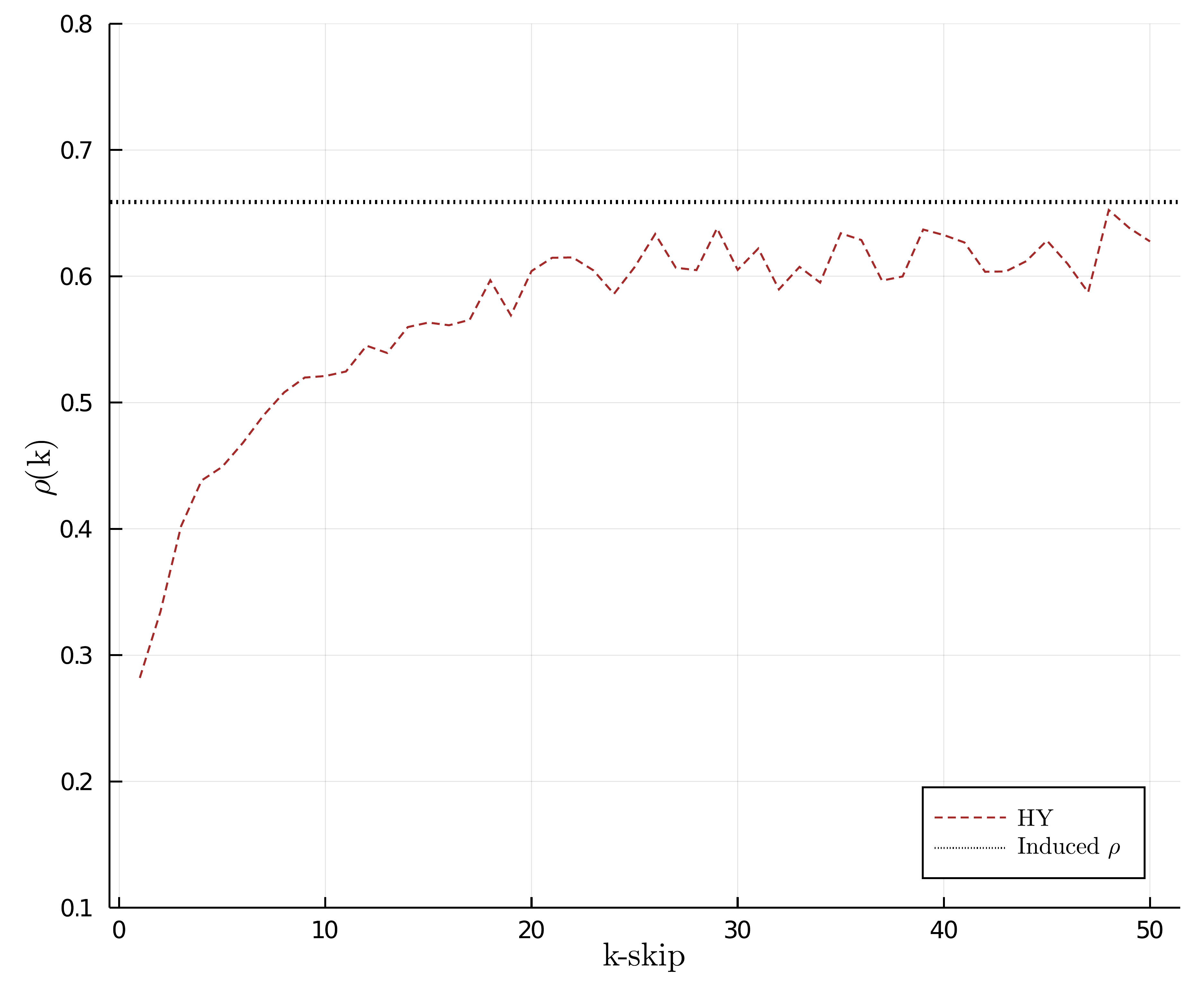}}~~
    \subfloat[Hawkes price model.]{\label{fig:k_skip:c}\includegraphics[width=0.33\textwidth]{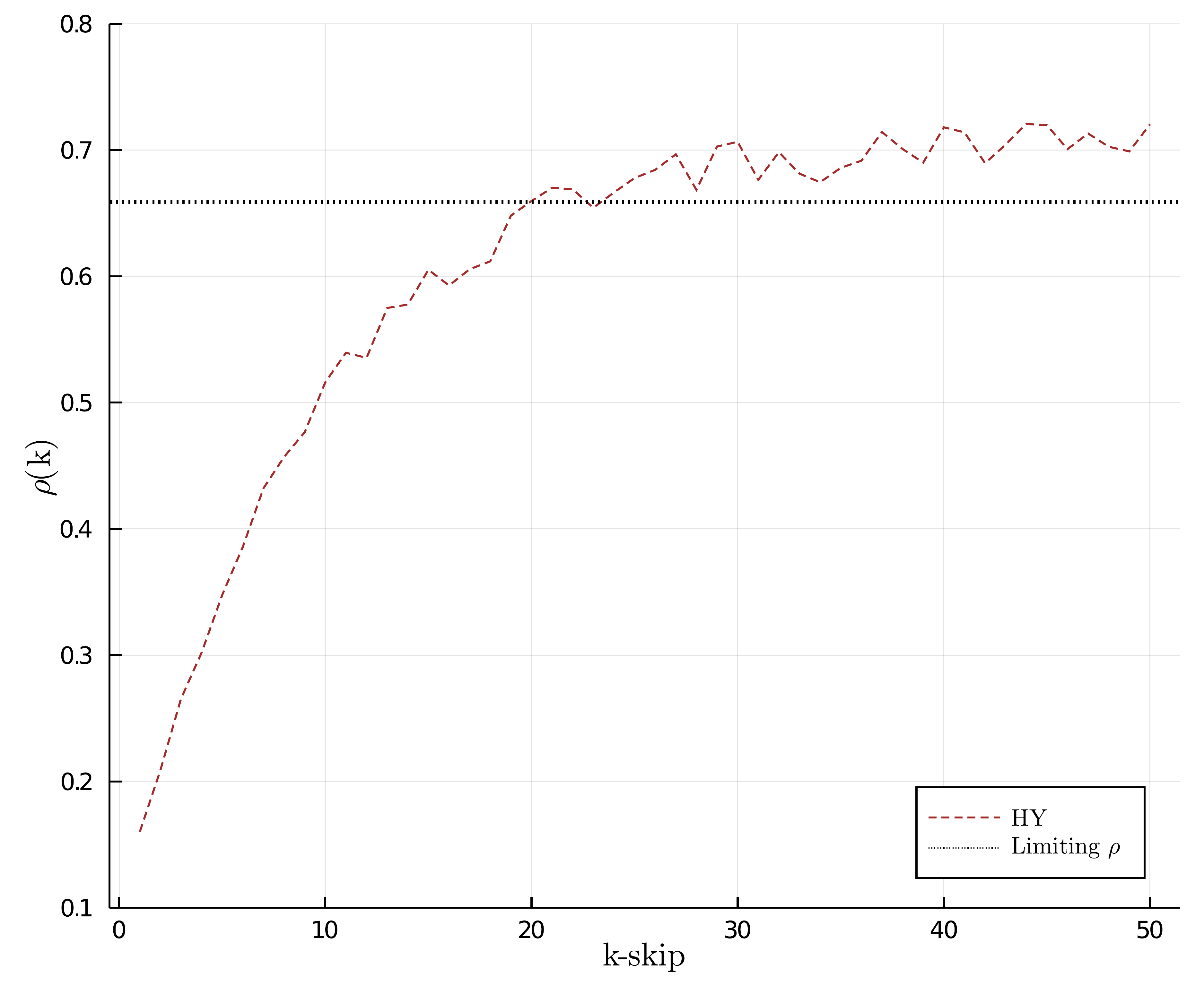}}
    \caption{The Hayashi--Yoshida estimate (dashes lines) given in Eq. \eqref{epps:eq:17} plotted as a function of the $k$-skip sampling for (a) the Brownian price model, (b) the MMN model and (c) the Hawkes price model. The limiting/induced $\rho$ (horizontal dotted line) is approximately 0.65 as before. We see that $k$-skip sampling allows us to discriminate the processes with one set of arrivals.}
\label{fig:k_skip}
\end{figure*}

The first two experiments are well suited to determine the underlying process of an unknown system in a simulation scenario. This is because the first two experiments require the re-sampling of the process to obtain different sets of observables $U^i$ and $U^j$ with different sized inter-arrivals. This is a problem with TAQ data because there is only one set of $U^i$ and $U^j$. Moreover, the second experiment requires multiple replications to get a good gauge on the error ribbons. Using results based on one replication with the second experiment cannot meaningfully determine the underlying process. Error bands are required to clearly see that the process consistently decays. The first experiment does not require replications to determine the underlying process. This is because the error ribbons are very narrow. This means that the estimates are either flat or they decay.

This experiment is a variant of the first experiment but overcomes the issue of requiring multiple sets of $U^i$ and $U^j$. This is achieved by using a single set of $U^i$ and $U^j$ when discriminating. \citet{GO2011} provide the inspiration to over come this issue by using $k$-skip sampling. With only one set of arrivals $U^i$ and $U^j$, we can imitate multiple sets of arrivals with different sized inter-arrivals by sampling every $k$th observation in the single set of arrivals. Concretely, from the single set of $U^i$, let the $k$th-skip set of samples be the set $U^i_k = \left\{ t^i_{k}, t^i_{2k}, ..., t^i_{\lfloor \#U^i / k \rfloor k} \right\}$. Note that $\#U^i$ is the cardinality of set $U^i$. Instead of re-sampling and computing the Hayashi--Yoshida estimate with different mean inter-arrival $1/\lambda$, we can now compute the Hayashi--Yoshida estimate as a function of the $k$th-skip set of samples $U^i_k$ and $U^j_k$.

Figure~\ref{fig:k_skip} plots the Hayashi--Yoshida estimates as a function of the $k$-skip sampling for (a) the Brownian price model, (b) the MMN model and (c) the Hawkes price model. The $k$-skip sampling is performed by sampling every observation $k=1$ to every $50$th observation $k=50$. To re-create conditions similar to that of empirical data, we only sample one set of observables from each price process. Figure~\ref{fig:k_skip} recovers similar results as Figure~\ref{fig:HawkesPriceModelHYSamplingFreq}. We see that the induced correlation of the Brownian price model is recovered for any sized $k$-skip sampling, whereas the MMN and Hawkes price model presents an Epps like effect as a function of the $k$-skip sampling.

This result allows us to discriminate the underlying process of an unknown system with one set of observables. If the plot is flat for various values of $k$ then the underlying process is a diffusion-based process. If the plot presents a decaying correlation for smaller $k$ then the underlying process is either a diffusion-based process but contaminated with {\it ad hoc} noise or is simply composed of discrete connected events.

\section{Estimators for market microstructure noise}\label{sec:NoiseEst}

\begin{figure*}[h]
    \centering
    \subfloat[Brownian price model]{\label{fig:SuppDem:a}\includegraphics[width=0.33\textwidth]{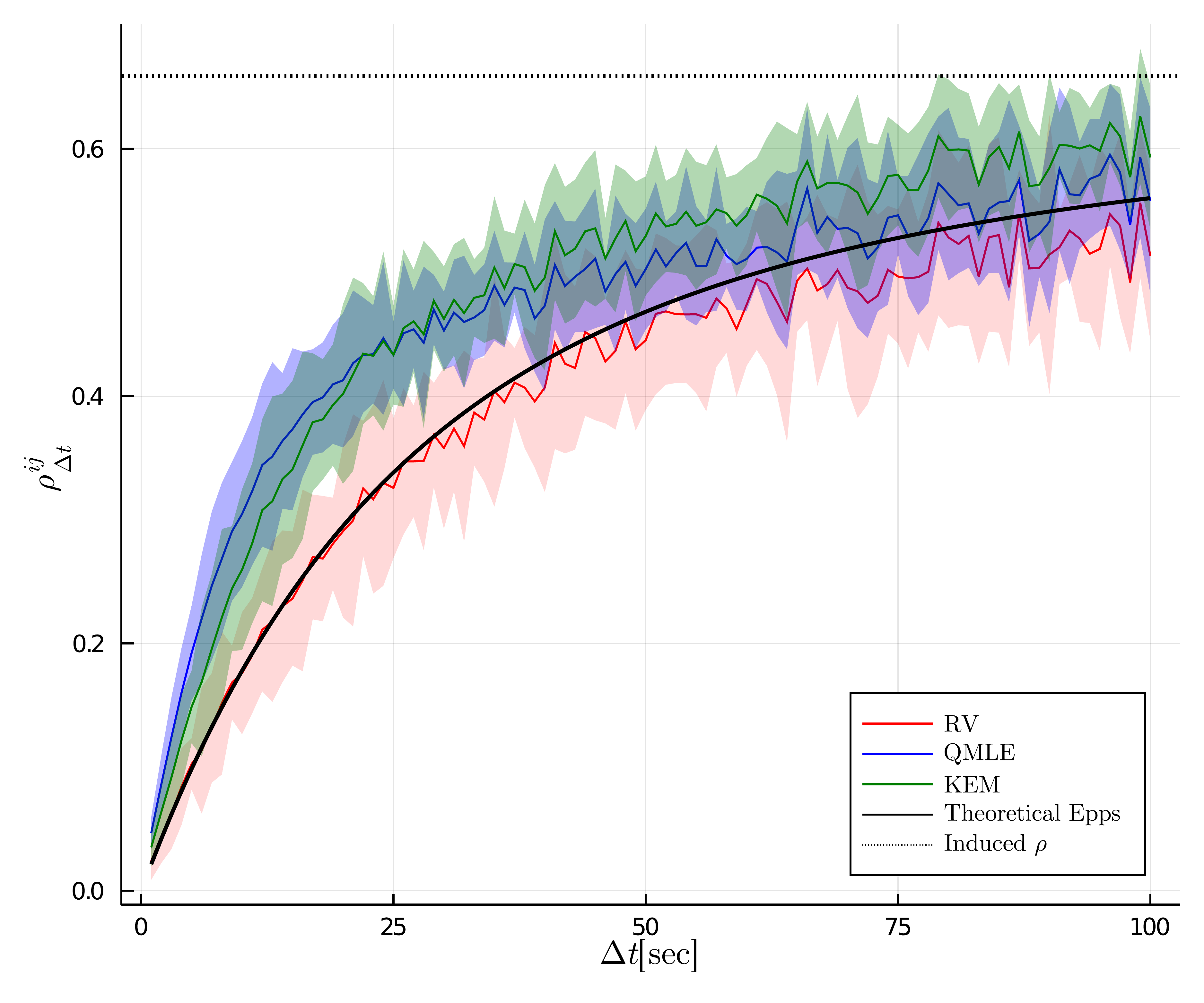}}
    \subfloat[MMN model]{\label{fig:SuppDem:b}\includegraphics[width=0.33\textwidth]{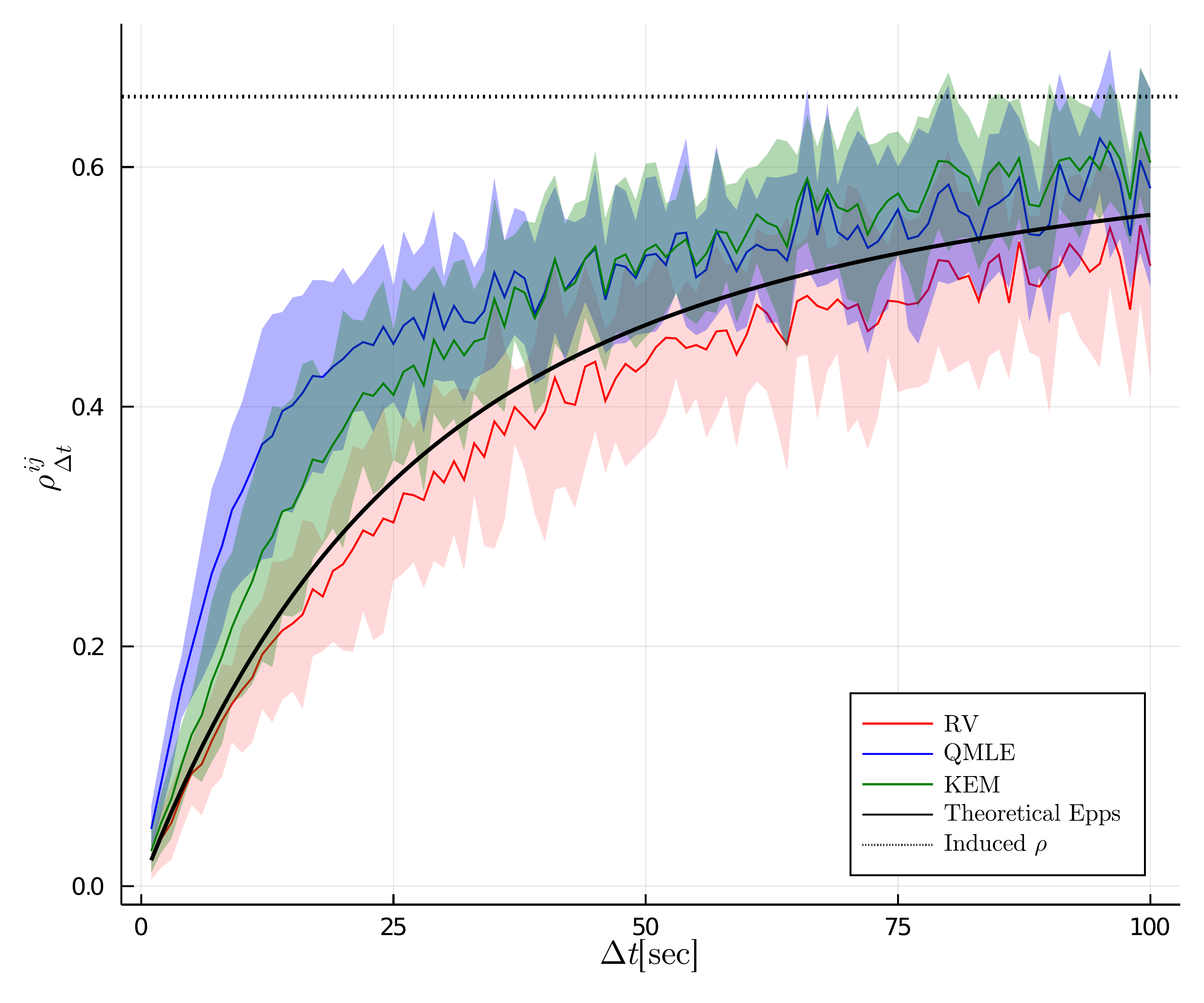}}
    \subfloat[Hawkes price model]{\label{fig:SuppDem:c}\includegraphics[width=0.33\textwidth]{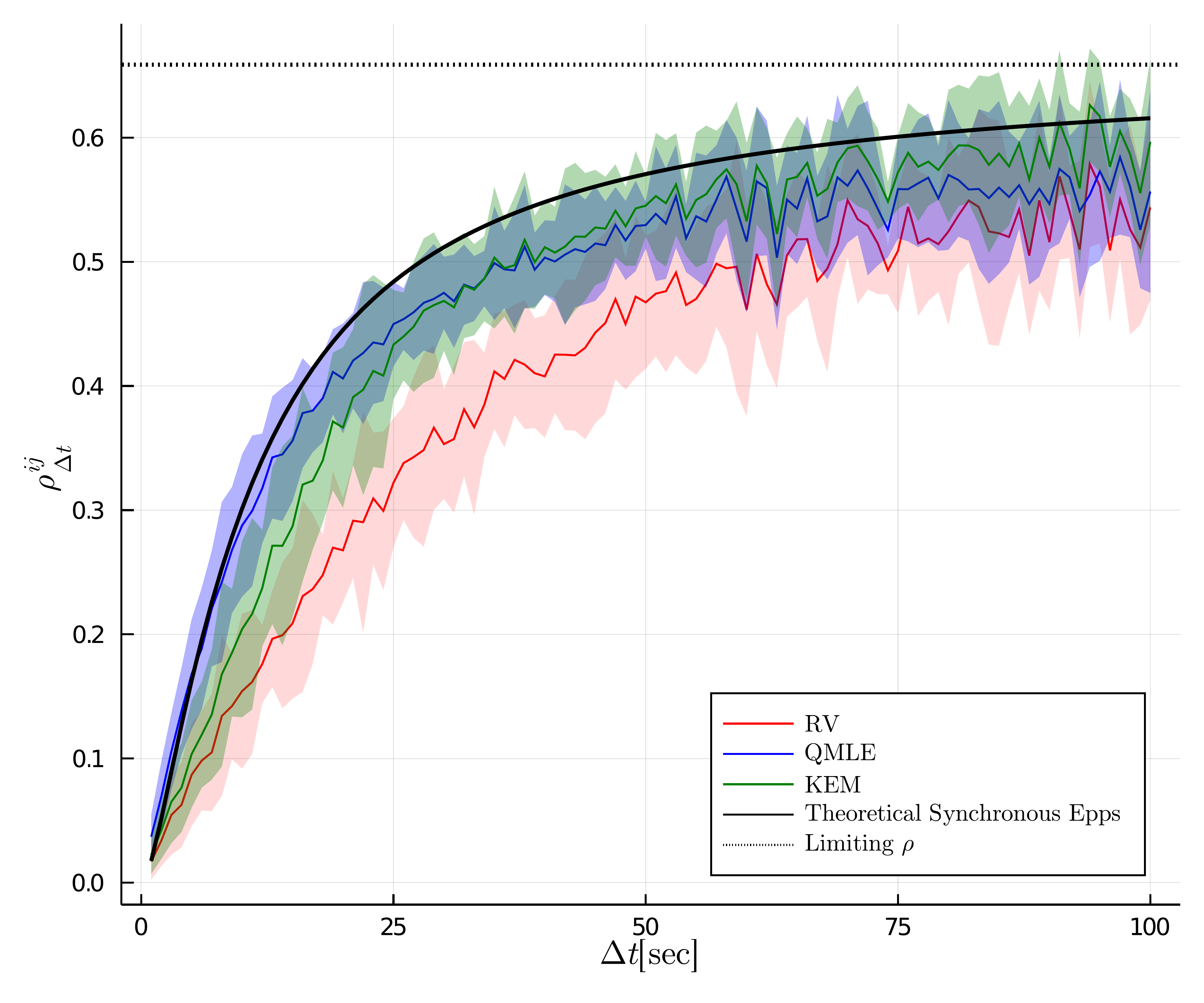}}
    \caption{The measured correlation (red line), QMLE (blue line) and KEM estimate (green line) are compared on (a) the Brownian price model, (b) the MMN model and (c) the Hawkes price model using Poisson sampling.}
\label{fig:SuppDem}
\end{figure*}

\begin{figure*}[htp]
    \centering
    \subfloat[QMLE on MMN model]{\label{fig:SuppExp2:a}\includegraphics[width=0.45\textwidth]{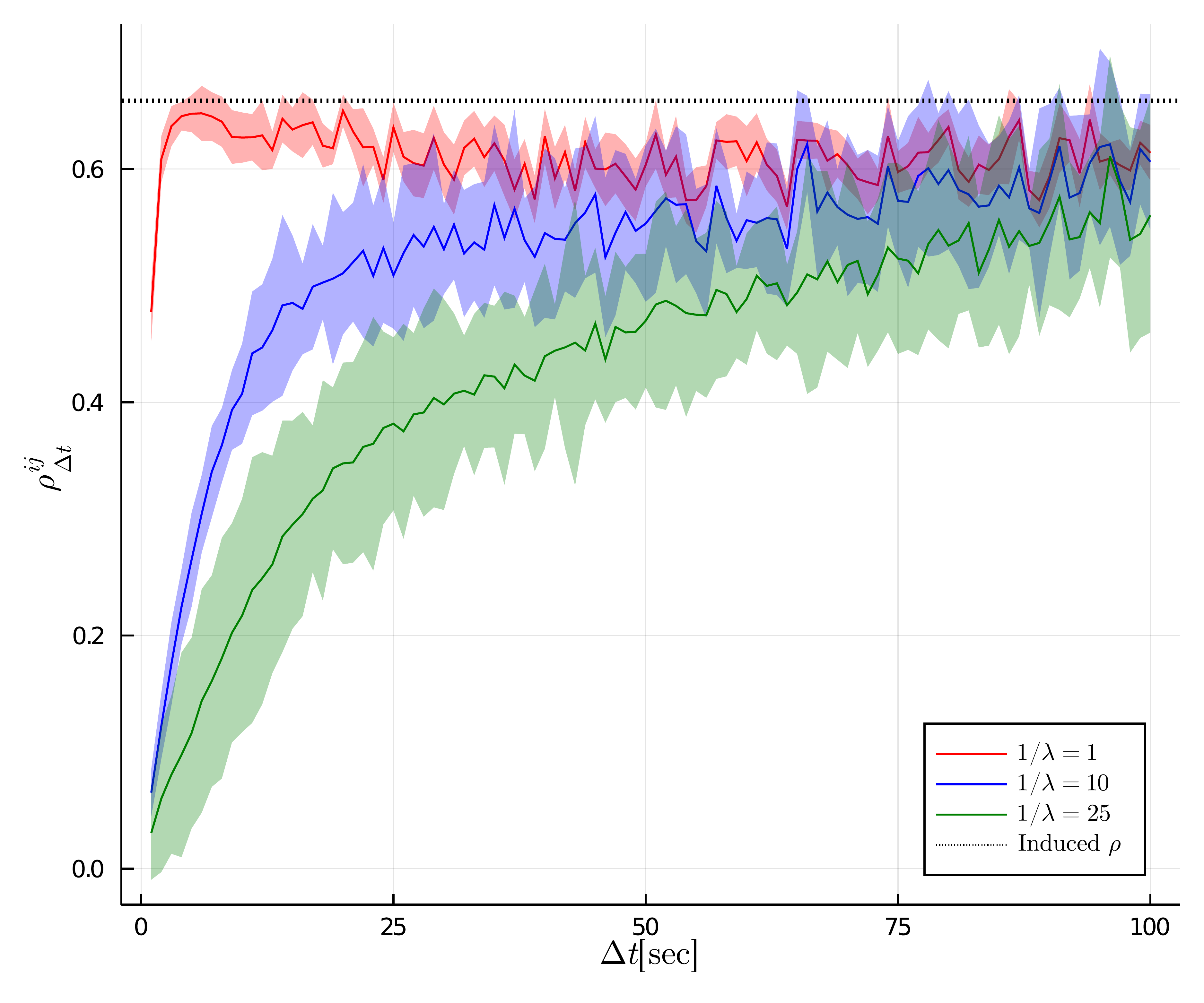}}
    \subfloat[QMLE on Hawkes price model]{\label{fig:SuppExp2:b}\includegraphics[width=0.45\textwidth]{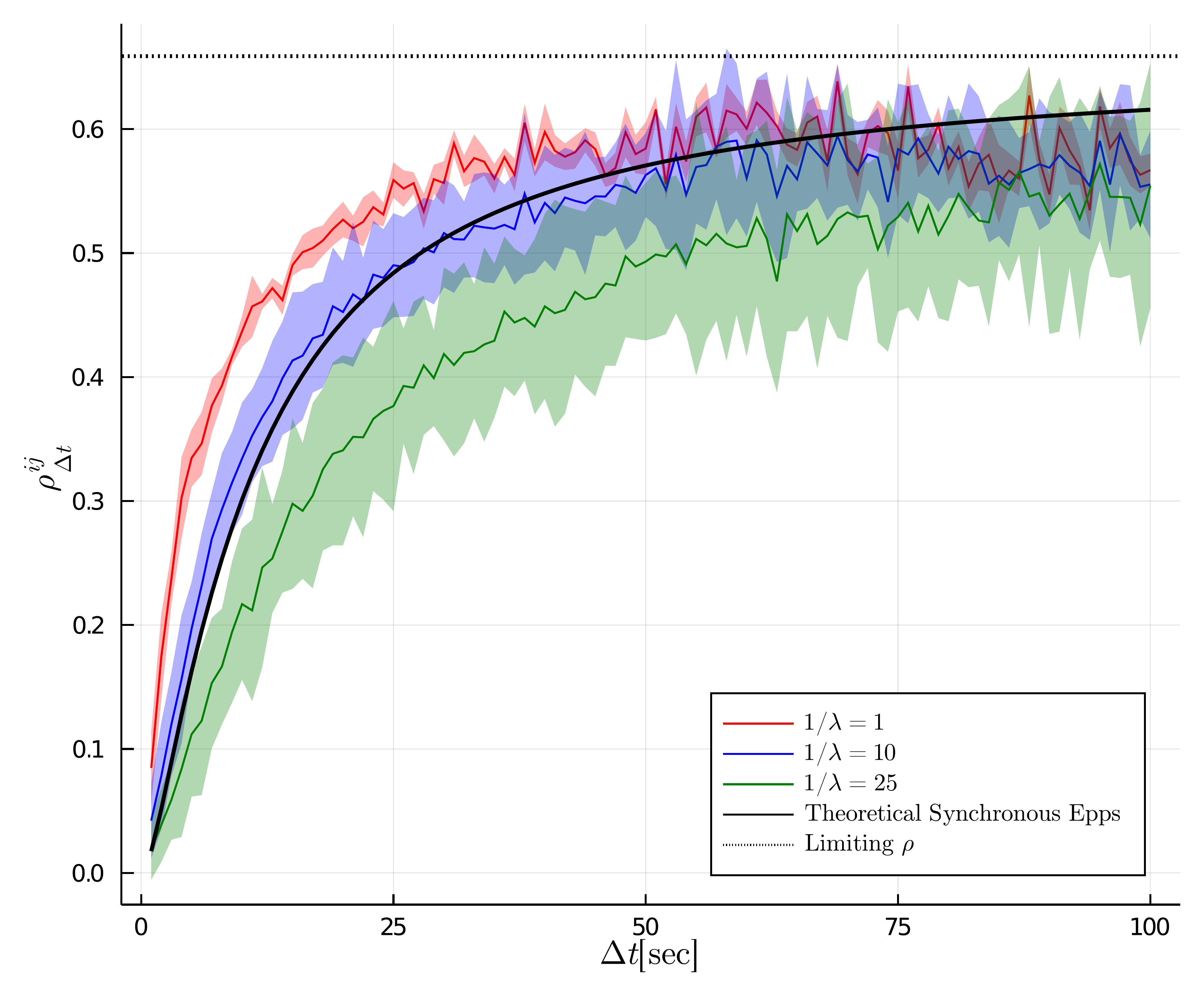}}    \\
    \subfloat[KEM on MMN model]{\label{fig:SuppExp2:c}\includegraphics[width=0.45\textwidth]{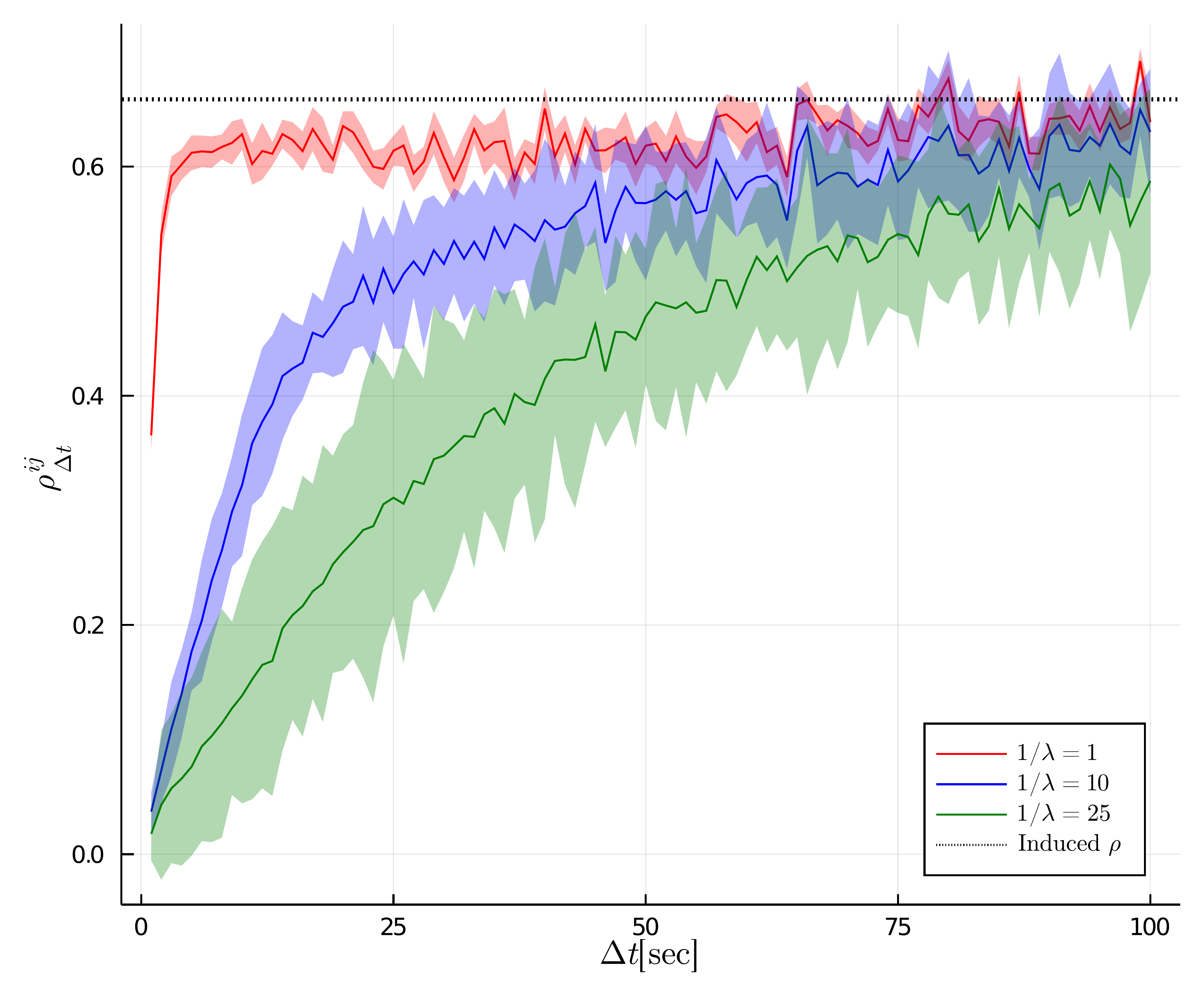}}
    \subfloat[KEM on Hawkes price model]{\label{fig:SuppExp2:d}\includegraphics[width=0.45\textwidth]{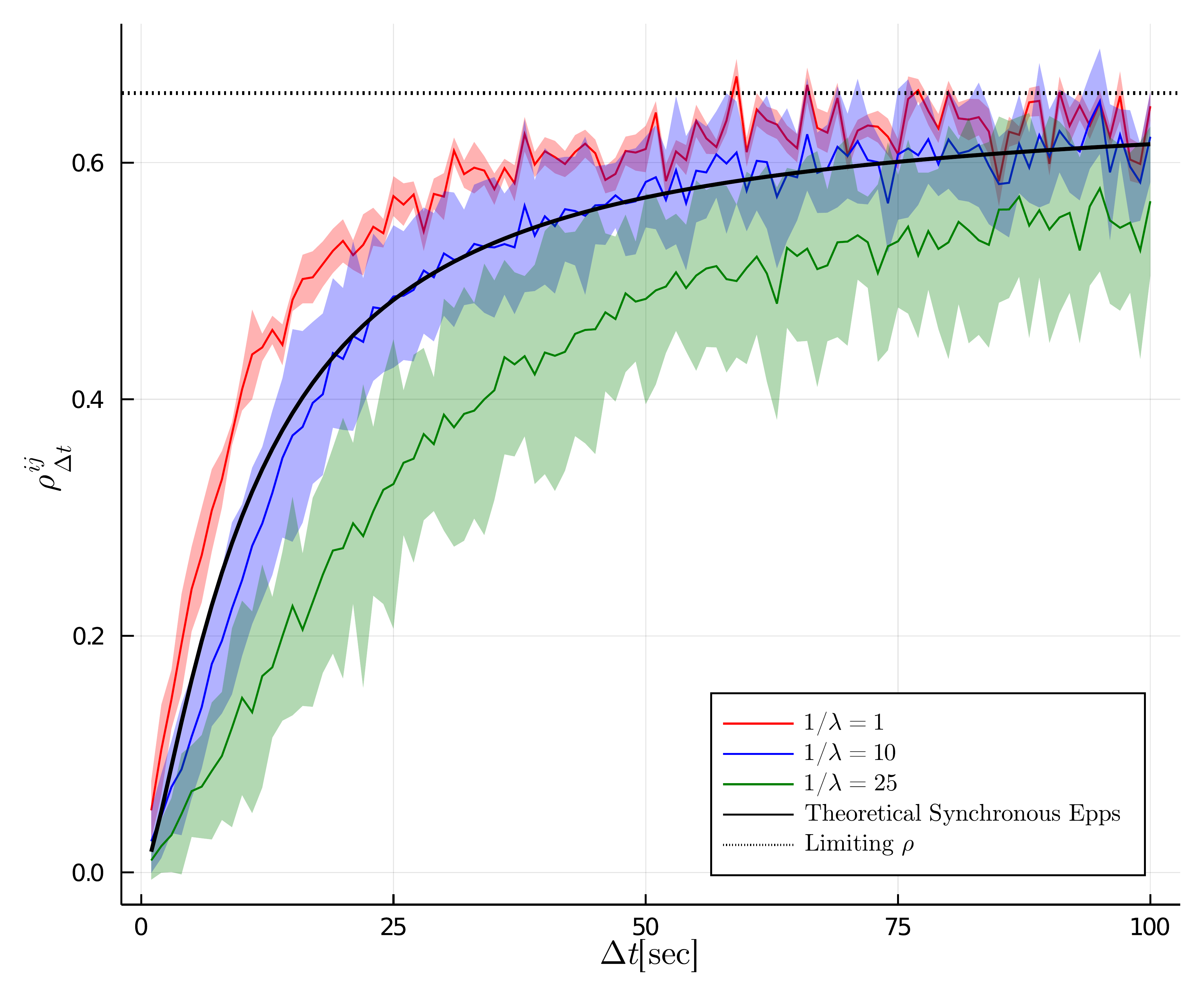}}
    \caption{The QMLE (first row) and KEM (second row) applied on the MMN (first column) and the Hawkes price model (second column) for three sampling frequencies of observables. The three sampling frequencies chosen are: $1/\lambda = 1, 10, 25$ which corresponds to the red, blue, green lines respectively. The thick line is the theoretical synchronous Epps effect given by Eq. \eqref{exp:eq:5} and the horizontal dotted line is the induced/limiting $\rho$ set to be approximately 0.65 as before.}
\label{fig:SuppExp2}
\end{figure*}

We apply two alternative estimators in order to try and discriminate between the MMN and Hawkes price model. The two estimators considered here are the Quasi-Maximum Likelihood Estimator (QMLE) \cite{AFX2010} and the Kalman-EM (KEM) estimator \cite{CPA2015}, which is a special case of the Multi-asset Lagged Adjustment (MLA) estimator 
\cite{BCP2020}.\footnote{We do not use the MLA estimator here because we found that the estimator overestimates the correlation in our models which have no lead-lag dependencies by trying to account for lead-lags.}

The QMLE and KEM estimator addresses both asynchrony and market microstructure noise. However, the way the two estimators address asynchrony does not fit easily within our experimental framework. 

First, the QMLE addresses asynchrony by applying a different sampling approach called the generalised sampling time. We use the most naive version of the sampling scheme that reduces to previous tick interpolation so that the estimator can fit within our experimental framework.

Second, the KEM addresses asynchrony by using the Kalman filter based smoother to access the latent states. The caveat here is that the observations and missing observations must all occur on a homogeneous grid. This is problematic within our experimental framework because we apply Poisson and Hawkes sampling to obtain the observables instead of a Bernoulli censoring approach. Even though Poisson sampling and independent Bernoulli censoring are equivalent, the subtle difference is that observables from Poisson sampling do not necessarily lie on a homogeneous grid. This creates an ambiguity when creating a homogeneous grid at different time scales $\Delta t$. For small $\Delta t$, if an observable is near two grid points, a decision needs to be made in terms of which grid point the observable should be shifted to. For large $\Delta t$, if there are more than two observables near the same grid point, a decision needs to be made in terms of which observable should be on that grid point. To overcome these issues, we simply apply previous tick interpolation and create a synchronised grid to feed into the KEM estimator. This also ensures that the estimator fits within our experimental framework.

We reduce the simulation size to $T$ being 6.5 hours (23,400 seconds) and we only perform 10 replications for this proof of concept. The reason for this is because of the long compute time of the QMLE which requires the optimisation of four likelihoods to obtain a correlation estimate. Otherwise the parameter settings for the models and the experimental setup remain the same as before.

Figure~\ref{fig:SuppDem} compares the measured correlation, QMLE and KEM estimates for the (a) the Brownian price model, (b) the MMN model and (c) the Hawkes price model under Poisson sampling. We see that the QMLE and KEM estimates are larger than that of the measured correlation for all three models. This larger estimate is a result of the QMLE and KEM adjusting for market microstructure noise which may or may not be present depending on the model.

We perform experiment 2 from Section~\ref{sec:key} using the two estimators to see if they can help us discriminate between the MMN and the Hawkes price model. This can be seen in Figure~\ref{fig:SuppExp2}. We see that both processes lead to similar behaviours in the residual Epps effect. The case when $1/\lambda=1$ is notable as these arguably present different behaviours between the two models. However the difference is not as clear cut as in the case of a simple Brownian motion against these two processes. Therefore it does not seem like discrimination between the two processes is possible with our current experiments.

What is interesting about Figure~\ref{fig:SuppExp2} is that the behaviour of the residual Epps effect under different sampling frequencies (from the observables) are the reversal of what is seen in Figure~\ref{fig:HawkesPriceModelwDiffSamplingFreq}. We see that there is a larger (smaller) decay in the residual Epps effect for larger (smaller) $1/\lambda$. We argue that this is a result of the interplay between the estimators correcting for noise and the signal-to-noise ratio from the different time scales induced by the observables. The estimators pick up a larger signal-to-noise ratio which leads to a larger correction when $1/\lambda$ is small, and the estimators pick up a smaller signal-to-noise ratio which leads to a smaller correction when $1/\lambda$ is large. 

What is concerning is that the estimators also seem to be correcting for noise in the Hawkes price model. The estimators are misconstruing the fundamental discreteness in the model as noise. This is because the noise assumption is {\it a-priori} built into the estimators. Addressing this {\it a-priori} assumption should be considered when picking estimators and designing experiments to discriminate between the two processes. 
Other estimators that could potentially help us discriminate between the MMN and the Hawkes price model include the estimators from \cite{BHLS2011,PCM2014} and \cite{MSG2011}. Additionally, the MLA estimator from \cite{BCP2020} could prove useful when trying to discriminate under the presence of lead-lag.

\section{Perspectives on noise and discreteness}\label{sec:noise}

\subsection*{Limitations of the discrimination}

In Section~\ref{sec:key} we were able to discriminate between a diffusion-based process and a diffusion-based processes with {\it ad hoc} noise, or discrete connected events. However, we cannot discriminate between diffusion-based processes with {\it ad hoc} noise and discrete connected events. This is because both of these processes lead to similar behaviour in the residual Epps effect, albeit for completely different reasons.

In Section~\ref{sec:NoiseEst} we applied two alternative estimators to demonstrate the complexity and ambiguity relating to attempts to discriminate between the MMN and Hawkes price model using the second experiment. Specifically, we applied the Quasi-Maximum Likelihood Estimator (QMLE) \cite{AFX2010} and the Kalman-EM (KEM) estimator \cite{CPA2015} to the problem. We find that both processes lead to similar behaviours in the residual Epps effect and it remains unclear on how we can discriminate between these processes.

Another possible limitation of our experiments is that they are not designed to account for lead-lag. This could be problematic because \citet{GO2011} recovered behaviour similar to what was observed in Figures~\ref{fig:k_skip:b} and \ref{fig:k_skip:c} using empirical TAQ data from the New York Stock Exchange (NYSE) with the $k$-skip Hayashi--Yoshida estimator. However, they argue that price dynamics are slow to fully reflect the appropriate correlations in the empirical data. The underlying assumption of the Hayashi--Yoshida estimator is that all available information regarding the correlation should be fully incorporated when a price update arrives. They then correct for this by adjusting the Hayashi--Yoshida estimator to account lead-lag dependencies.

Lead-lags can arise within the framework of the Hawkes price model \cite{BDHM2013a, BDHM2013b}, however the specification used here in Eq. \eqref{exp:eq:4} is fully symmetric and thus there are no lead-lag effects in our models (see Remark 9 of \citet{BDHM2013b}). Currently our discrimination is performed without considering lead-lag effects but this is something that could be incorporated when designing future experiments but does not undermine our claims. 


\subsection*{Finding the correct representation}

Intraday financial prices are subject to discretisation effects where the prices are multiples of the tick-size \cite{RobertChristianY.2011Anaf}. This is a fundamental property arising from how financial markets are designed. This property is naturally incorporated within tick-by-tick models such as the Hawkes price model. In contrast, this property poses a problem for macroscopic models such as those built using Brownian motion (diffusion models) because the implied resolution dynamics under such models are at odds with the underlying process --- something which becomes immediately clear when estimating correlation structures \cite{MSG2011,Robert2011}.

The ``market microstructure noise effects" were introduced to reconcile macroscopic latent models with the signature plot \cite{AMZ2005,AMZ2011,ZMA2005}. That is, by incorporating noise, various microstructure effects due to the discreteness of price dynamics {\it e.g.} bid-ask bounces and measurement or rounding errors can be accounted for. However, we argue that although with enough such additions one can reconcile empirical effects with a chosen latent model they do not reconcile the model process with the data generating process at a microscopic resolution. For this reason we advocate that noise is technically an additional {\it ad hoc}\footnote{\citet{Popper1974}, pp. 986, “I call a conjecture “ad hoc” if it is introduced […] to explain a particular difficulty, but if […] it cannot be tested independently”; here it may be best to think about the noise assumption as ``unwarranted" following \citet{Hunt2012}.} modelling assumption that is used to reconcile the discrepancy between tick data and macroscopic models when ``downscaling'' from macroscopic to microscopic \cite{AJ2020}.

Finding a more appropriate representation for the tick data generating process is particularly important when it comes to the investigation of the Epps effect. This is because the decay in correlation is an anomaly that requires a correction under the noise representation. The goal of the noise representation is to try recover the constant underlying induced correlation that is independent of $\Delta t$ under the presence of {\it ad hoc} noise.

On the other hand, the decay in correlation is a fundamental property under the representation of discrete connected events. The ambition here is then to only correct for statistical effects leading to a decay in correlations so that we can recover the theoretical underlying correlation that does depend on $\Delta t$. Indeed, starting with a representation that recovers fundamental properties at the atomic scale may still resolve effects when transitioning to macroscopic observation scales, {\it e.g.} \citet{DIANA2025116202} provide a discrete model that can be averaged to generate the emergence of an Epps effect \cite{bauer2024correlation}. In this model the system comprises discrete events that do not naturally align in a unique global trading time and there are events that cannot be uniquely aligned in a cross sectional epoch. Correlations can only emerge on averaging scales sufficiently long that there are enough events in the discrete sample periods to allow meaningful estimation. Correlations are then an extrinsic property as the result of an averaging procedure that is separate from the data generating process.

The key point we are making is that additional {\it ad hoc} model complexity is required in order to recover the observed phenomena from models which are based on Brownian motions. This type of model complexity is not required when using a Hawkes representation which starts with the event waiting time distributions, nor is it required when starting with order-flow models built on market phenomenology \cite{DIANA2025116202} that start with discrete order events. We argue, based on the principle of parsimony, {\it i.e. Occam's razor}, that the Hawkes representation is a more appropriate foundational stochastic process based representation, and that the simplest explanation for the Epps effect is that it is the signature, the ``smoking gun'', of a fundamentally discrete and event driven data generating process.

Indeed, the discrete nature of real-world market processes do not only arise through the lens of correlation. Discreteness at the atomic scale shows up even when attempting to model individual trajectories using jump diffusion models; Detectability of jumps in such processes using discretely observed trajectories is heavily dependant on the resolution (both length and frequency of the series) of the data\footnote{See Chapter 4 of \citet{pienaar2016non}}. 

More importantly, the extent to which one can distinguish between pure diffusion, diffusion with jumps, and pure jump processes such as Hawkes processes becomes similarly dependant on the observation resolution -- note all three of these model classes can be subsumed within the jump diffusion class\footnote{For example, see Figure 4.8.3 of \citet{pienaar2016non} and compare the transitional density dynamics of other jump diffusion models in the text, and note that these transitional densities are the lens through which we make inference at finite time scales.}. However, careful scrutiny of how such competing models behave under various resolutions has made it clear to the present authors that the diffusion component, at least as far as it concerns financial data, is a mathematical convenience that allows one to model at finite scales without the awkwardness of measuring events that occur at frequencies which cannot be observed at the time scale of the observations, and such discrepancies point to the underlying data generating process being discrete. When modelling at the atomic scale, diffusion-based models, direct or latent do not reflect the dynamics of the true data generating process.   

\section{Empirical Investigation}\label{sec:emp}

We compare the correction methods and detect the underlying processes using banking equities from the Johannesburg Stock Exchange (JSE) with TAQ data for 40 trading days starting from 2019/05/02 to 2019/06/28. The data is extracted from Bloomberg Pro and is processed so that trades with the same time stamp are aggregated using a volume weighted average. The equities considered are: FirstRand Limited (FSR), Absa Group Ltd (ABG), Nedbank Group Ltd (NED) and Standard Bank Group Ltd (SBK).

The correlations are measured for each trading day and the ensemble is reported. The error ribbons are computed such that they contain 95\% of the estimates at each $\Delta t$ between the trading days. They are computed using the Student $t$-distribution with 39 degrees of freedom and the standard deviation of the estimates between the days at each $\Delta t$. Additionally, $t=0$ starts once the equity pair has each made their first trade and $T = 28,200$ seconds for a seven hour and 50 minute trading day.

\begin{table}[htb]
\centering
\caption{The table reports the mean inter-arrival estimate and the associated standard deviation for the four banking equities over the 40 day trading period.}
\begin{tabular}{ccc}
  \hline
Tickers & $1/\hat{\lambda}$[sec] & $\hat{\sigma}(1/\lambda)$ \\ 
  \hline
FSR & 12.09 & 19.53 \\ 
SBK & 13.06 & 21.51 \\ 
NED & 15.39 & 25.14 \\ 
ABG & 15.68 & 26.27 \\ 
  \hline
\end{tabular}
\label{tab:data}
\end{table}

\begin{figure*}[h]
    \centering
    \subfloat[SBK/FSR.]{\label{fig:empcorrect:a}\includegraphics[width=0.5\textwidth]{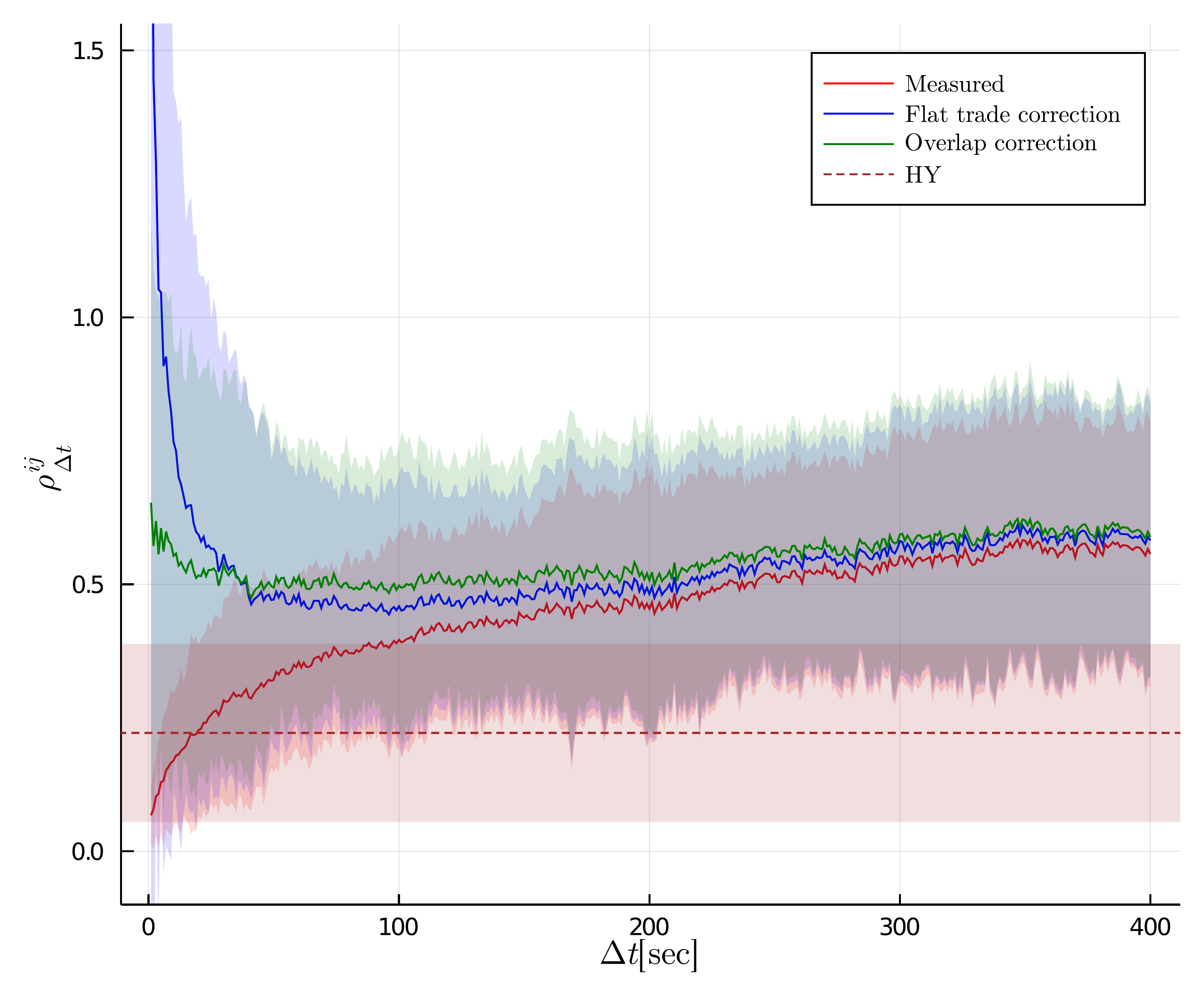}}
    \subfloat[NED/ABG.]{\label{fig:empcorrect:b}\includegraphics[width=0.5\textwidth]{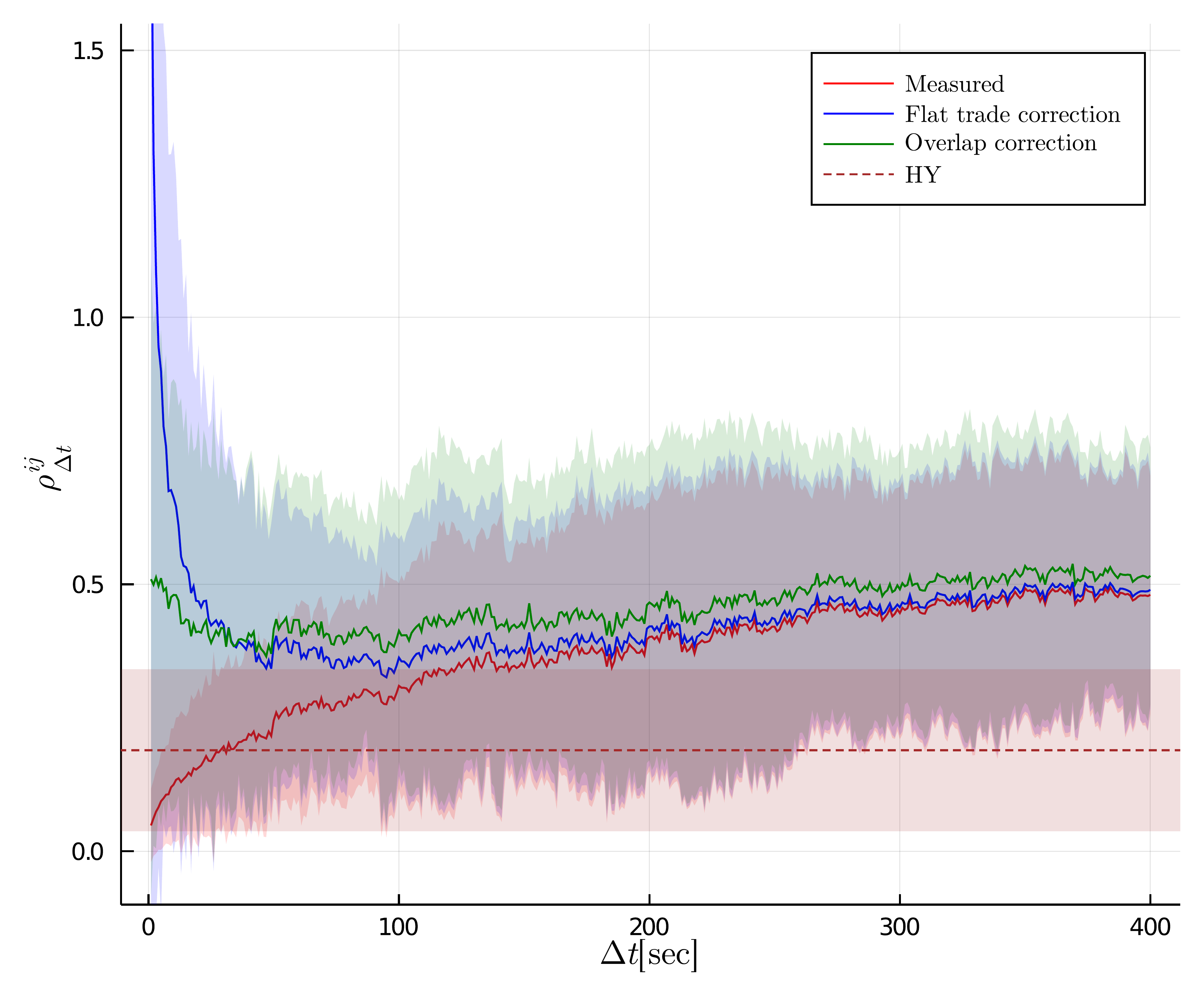}}
    \caption{The corrections are applied and compared for the equity pairs (a) SBK/FSR, and (b) NED/ABG. The red line is the measured correlation from the synchronised process Eq. \eqref{epps:eq:measured}, the blue line is the flat trade correction Eq. \eqref{epps:eq:14} and the green line is the overlap correction Eq. \eqref{epps:eq:13}. Lastly the horizontal dashed line is the Hayashi--Yoshida estimator Eq. \eqref{epps:eq:17}.}
\label{fig:empcorrect}
\end{figure*}

\begin{figure}[htb]
    \centering
    \includegraphics[width=0.48\textwidth]{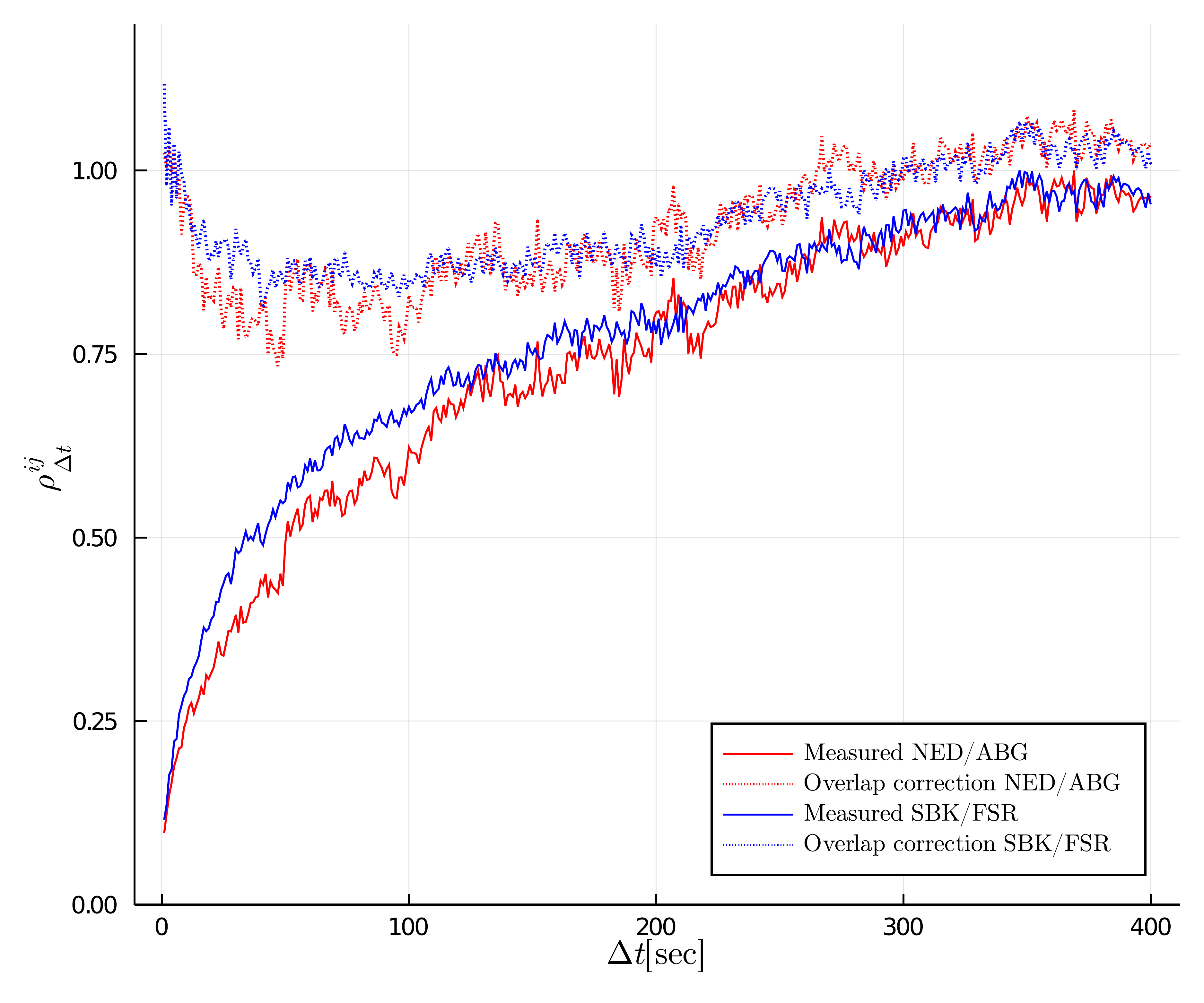}
    \caption{Comparison of the overlap correction (dotted lines) and the measured correlation (solid lines). All the estimates are scaled by the saturation correlation so that clearer comparisons can be made for the equity pairs NED/ABG (red) and SBK/FSR (blue).}
    \label{fig:EmpComparison}
\end{figure}

\begin{figure}[htb]
    \centering
    \includegraphics[width=0.48\textwidth]{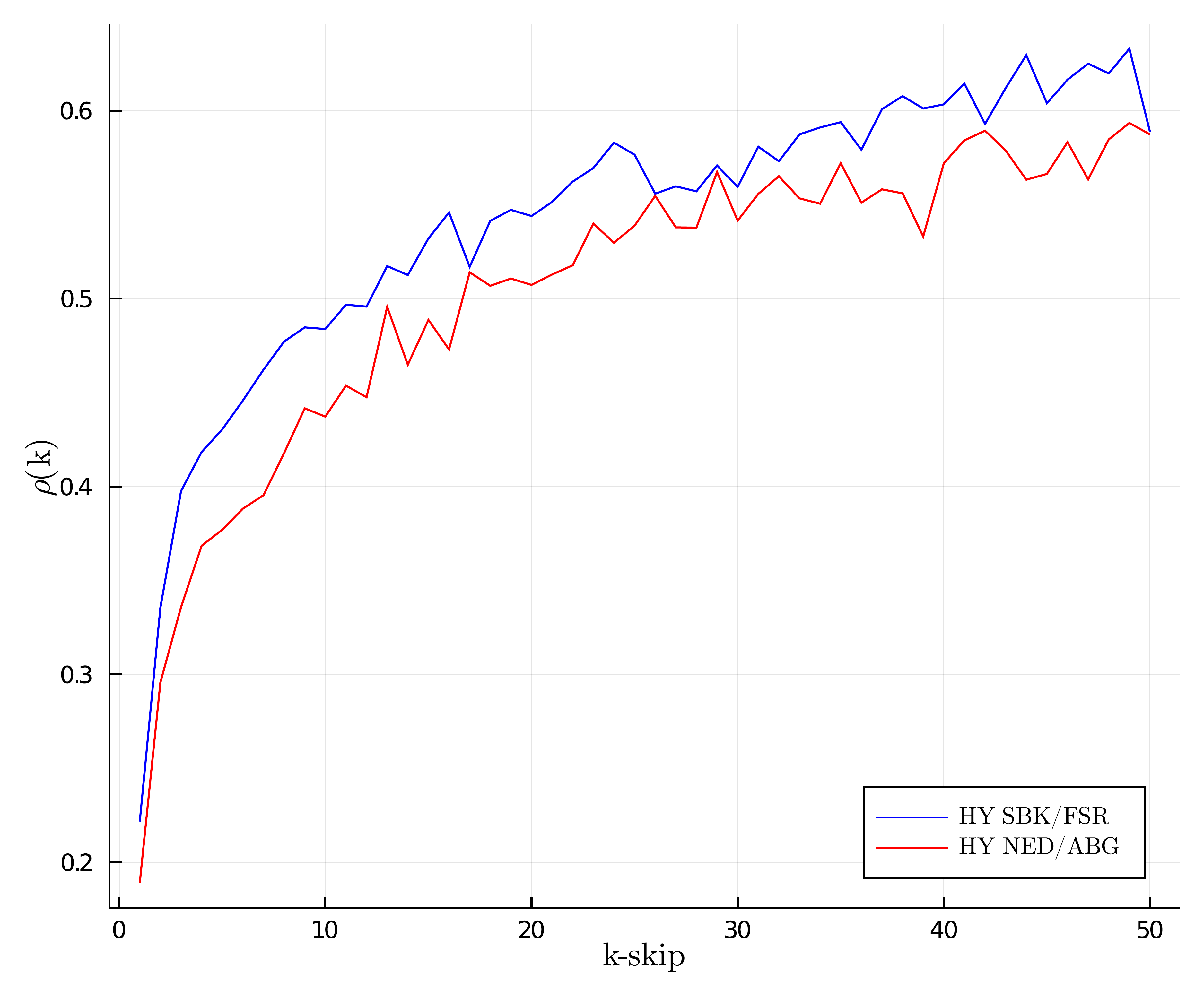}
    \caption{The $k$-skip sampling with the Hayashi--Yoshida estimator applied to the equity pairs SBK/FSR (blue line) and NED/ABG (red line). The empirical reality suggests the underlying system is from discrete connected events.}
    \label{fig:Empkskip}
\end{figure}

We only have one set of observables from empirical TAQ data which means that the second experiment is not well suited for discriminating empirical data. Nonetheless we attempt to apply the experiment because it does reveal something interesting. In order to apply the second experiment we try to re-create the simulation setting by finding equity pairs with similar correlation levels but different average inter-arrivals. Table~\ref{tab:data} gives the estimate for the mean and standard deviation for the inter-arrival time over the 40 trading day period. It is clear that the pair FSR and SBK (NED and ABG) have comparable and on average smaller (larger) inter-arrival times respectively. Therefore we will compare the corrections on FSR/SBK against NED/ABG to determine if there is a difference in the decay of the residual Epps effect following the spirit of the second experiment.

Figure~\ref{fig:empcorrect} compares the correction methods on the pair (a) FSR/SBK and (b) NED/ABG to investigate the residual Epps effect. We see that the variability of the correlation estimates between days is very high but largely remains positive. This is why we chose the banking equities since they have a strong correlation. If the equities are less correlated, the Epps curves can jump between negative and positive between the days. The flat trade correction again over-compensates the correction. The Hayashi--Yoshida estimates do not recover the same estimates as the overlap correction or the saturation level. This is an indication that empirical TAQ data cannot be modelled using Brownian motions alone.

Figure~\ref{fig:EmpComparison} takes a closer look at the overlap correction and the measured correlation for the equity pairs SBK/FSR and NED/ABG investigated in Figure~\ref{fig:empcorrect}. Here we re-scale the estimates by the saturation level for better comparison since we have two different equity pairs. We see the measured and correction are similar for both the equity pairs. The similarity of the overlap correction may be due to the fact that the average inter-arrivals are not significantly different. What is more interesting is this U-shaped residual Epps curve which has not been reported in the broader literature. This might be because the corrections are usually not applied at such a fine resolution. Here when $\Delta t > 100$ seconds the decay is present, the decay starts to flatten out for $\Delta t$ between 50 and 100 seconds, and below 50 seconds the residual correlation starts to rise back towards the saturation level. This behaviour could not be replicated in the simulations with the models and sampling methods considered. 

We apply the $k$-skip Hayashi--Yoshida to better discriminate the underlying process. Figure~\ref{fig:Empkskip} computes the Hayashi--Yoshida estimates with the $k$-skip sampling for $k$ ranging from 1 to 50 on the equity pairs SBK/FSR and NED/ABG. Both equity pairs present a decaying correlation for smaller $k$. This is strong evidence suggesting that empirical TAQ data cannot simply be modelled using a pure diffusion-based process. Additional model complexity such as market microstructure noise or a more appropriate representation using discrete connected events is required.

Even though the evidence in Figure~\ref{fig:Empkskip} may seem convincing, we have yet to control for the effect of lead-lag. Additional work is required to design experiments that incorporate the impact of lead-lags when discriminating the underlying process. This extension may possibly provide insight into this U-shaped residual Epps curve.

\section{Conclusion}\label{sec:conclusion}

In empirical market studies it can be argued the that fundamental feature of real markets are that they are discrete and event based. This then suggests that any stochastic process based representation should start with the waiting times between events and then motivate other averaged features from this. This provides an ambiguity in the representation if an underlying continuous-time representation is to be retained. To study this ambiguity we consider the Epps effect arising from a Hawkes process representation. 

We derive the Epps effect arising from asynchrony and provide a refined method based on existing work \cite{MSG2010,MSG2011} to correct for this effect. This has been informed by the rich prior literature \cite{BDHM2013a,BDHM2013b,MMZ2011,TK2007,TK2009,PI2007}. The refinement is simpler to compute as it separates the estimation of correlation from the correction factor without making strong assumptions about the distributional properties of the sampling processes. 

The method is compared against the Hayashi--Yoshida estimator \cite{HY2005} and a correction using the probability of flat trading \cite{BLPP2019} on three models. We found that our correction recovers similar estimates to the Hayashi--Yoshida estimator and outperforms the flat trade correction.

We design three experiments to discriminate the underlying process between diffusion-based processes against diffusion-based processes with {\it ad hoc} noise, and a process made from discrete connected events. The first two experiments are better suited for simulation scenarios because they repeatedly re-sample the process. The third experiment uses $k$-skip sampling and allows us to detect the underlying process with one set of observables.

These experiments demonstrate that discrimination is in principle possible but highlights the complexity relating to discriminating between diffusion-based processes with {\it ad hoc} noise against discrete connected events. This is because the two processes recover similar correlation dynamics albeit for very different reasons. We argue that the Hawkes representation is the more appropriate representation because it is the simpler model that recovers the observed phenomenology without {\it ad hoc} assumptions.

The experiments are applied to trade and quote data from the Johannesburg Stock Exchange. We found an interesting U-shaped residual Epps curve which could not be replicated under simulation. We find evidence using the $k$-skip sampling suggesting that diffusion-based process alone are insufficient to model empirical trade and quote data. The modelling should be done using discrete connected events or the diffusion process needs to be contaminated with {\it ad hoc} noise.

If the Epps effect is not just a statistical anomaly that requires a correction then this would mean that we need to be more careful when considering the impact of time scales in decision making if the decisions are predicated on estimates of correlation matrices. In particular, investigating the financial implications and finding the optimal scale(s) to estimate correlation matrices for different types of financial decisions will be a useful exercise.

If tick-by-tick models such as the Hawkes processes can point to an appropriate stochastic process representations then a reconciliation of time-series properties across various time scales is necessary. Nearly unstable Hawkes processes \cite{JR2015a,JR2015b} can be seen as the first steps towards a unified model that fits both microstructure stylised facts and diffusive stylised facts \cite{BMM2015} while recovering rough volatility \cite{GJR2018} and hence aspects of long-memory---all from the perspective of the fundamental nature of discrete data generating processes.

\section*{Acknowledgements}
The authors would like to thank colleagues and others in the community for constructive feedback. P.C. would like to acknowledge the support by the Manuel \& Luby Washkansky Scholarship and the South African Statistical Association [grant number 127931].

\bibliographystyle{abbrvnat}
\bibliography{PCEPTG-FMF}

\end{document}